\documentclass[12pt]{article}
\usepackage{mathrsfs}
\usepackage{natbib,graphicx,setspace,lscape,longtable}
\usepackage{mathrsfs,amsmath,amsthm,amssymb,color}
\usepackage{natbib,epsfig,graphicx,pdfpages}
\usepackage{hyperref}
\usepackage{rotating}
\usepackage{float}
\usepackage{bm}
\usepackage{xr}
\usepackage{ulem}
\usepackage{CJK}
\usepackage{natbib}
\usepackage{algorithm}
\usepackage{algorithmic}
\usepackage{booktabs}
\usepackage{bm}
\usepackage{subfigure}
\usepackage{makecell}
\usepackage{bbm}
\usepackage{xcolor}
\usepackage{bbm}
\usepackage{multirow}
\usepackage{textcomp}
\allowdisplaybreaks[2]
\bibpunct{(}{)}{;}{a}{,}{,}
\newtheorem{theorem}{Theorem}
\newtheorem{lemma}{Lemma}
\newtheorem{proposition}{Proposition}

\newcommand{\csection}[1]
{\begin{center}
\stepcounter{section}
{\bf\large\arabic{section}. #1}
\end{center}
}

\newcommand{\csubsection}[1]{
\begin{center}
\stepcounter{subsection}
{\it\arabic{section}.\arabic{subsection}. #1}
\end{center}
}

\hypersetup{
	colorlinks=true,
	linkcolor=blue,
	citecolor=blue,
	urlcolor=blue}

\def\n{\nonumber}

\def\beq{\begin{equation}}
\def\eeq{\end{equation}}
\def\beqr{\begin{eqnarray}}
\def\eeqr{\end{eqnarray}}
\def\beqrs{\begin{eqnarray*}}
\def\eeqrs{\end{eqnarray*}}
\def\bet{\begin{theorem}}
\def\eet{\end{theorem}}
\def\bel{\begin{lemma}}
\def\eel{\end{lemma}}
\def\bep{\begin{proposition}}
\def\eep{\end{proposition}}
\def\bg{\begin{figure}[tbph]\begin{center}}
\def\eg{\end{center}\end{figure}}

\def\bc{\begin{center}}
\def\ec{\end{center}}

\theoremstyle{definition}

\def\b0{\mathbf{0}}

\def\mR{\mathbb{R}}

\def\mX{\mathbb{X}}

\numberwithin{equation}{section}
\title{\Large\textbf{Popularity Regression-Based Latent Space Model for Large-Scale Sparse Network Analysis }}
\author{Zhan Gao$^1$, Danyang Huang$^{2,*}$, Rui Pan$^{3}$ and Hansheng Wang$^1$}
\date{$^1$ \textit{Guanghua School of Management, Peking University, Beijing, China}, $^2$ \textit{Center for Applied Statistics and School of Statistics, Renmin University of China, Beijing, China}, \\ $^3$ \textit{School of Statistics and Mathematics, Central University of Finance and Economics, Beijing, China}}
\begin{document}
\abovedisplayshortskip=4pt
\belowdisplayshortskip=8pt
\abovedisplayskip=4pt
\belowdisplayskip=8pt
\maketitle
\vspace{-1cm}
\begin{singlespace}
\begin{abstract}
Degree heterogeneity is one of the most important properties of network data. It is widely observed that degree heterogeneity is often related to the nodal features. In this study, we investigate the estimation and statistical inference for a popularity regression-based latent space model with nodal features. Given the complex dependence structure induced by the latent space model, we aim to derive analytically tractable objective functions that effectively account for this structure for sparse networks. Specifically, we propose a total of four estimators. The first two estimators are developed by utilizing only the first-order structure of the network (e.g., the nodal degree), while the last two estimators are developed by leveraging the higher-order network structures (i.e., reciprocity and transitivity). Rigorous asymptotic theory is established based on various non-standard $U$-statistics. We find that different estimators might have different convergence rates. The extension to higher-order moments-based estimators is also discussed. Extensive numerical experiments and a real data analysis of link prediction for an author citation network are conducted for illustration purposes.

\

\end{abstract}
\end{singlespace}
\newpage
\csection{INTRODUCTION}

Network data analysis has attracted great attention over the past decades. Unlike independent data, network data are inherently embedded with various connections among them, which in turn induces a sophisticated network dependence structure between nodes \citep{lee2004asymptotic}. Specifically, network data refers to the collection of nodes linked to relationships. Typical examples include social networks where nodes are users and links are friendship relationships \citep{hunter2008goodness}, co-authorship networks where nodes are authors and links are collaborations \citep{aoas2016Ji,ji2022co}, gene networks where nodes are molecular regulators and links are regulatory interactions \citep{chun2015gene}. To understand the underlying structure of networks, various models are proposed and widely used, including $p_1$ models \citep{holland1981exponential}, stochastic block models \citep{holland1983stochastic,wang1987stochastic}, latent space models \citep{hoff2002latent}, random dot product graph models \citep{young2007random}, and graph neural network models \citep{wu2020comprehensive}.

Among all these models, the latent space model (LSM) proposed by \cite{hoff2002latent} is one of the most extensively studied models due to its excellent explainability. A latent space model assumes that the nodes are embedded in an unobserved low-dimensional Euclidean space, and the probability of connection between two nodes increases as they are positioned closer to each other in this latent space. Important network structures such as reciprocity and transitivity can be naturally expressed by the latent space model. As a result, various extensions of the latent space model have been developed for complex networks, including dynamic networks \citep{sewell2015latent},
bipartite networks \citep{friel2016interlocking}, multilayer networks \citep{zhang2020flexible}, and multiplex networks \citep{macdonald2022latent}. Regarding the estimation of the latent space model, the Bayesian approach has been widely used \citep{hoff2002latent,sewell2015latent}. However, these approaches require Markov chain Monte Carlo type algorithms and become practically infeasible for large-scale networks, leading to a prohibitive computational burden. To alleviate this problem, the variational Bayesian method \citep{liu2022variational,zhao2024structured} and the projection gradient descent algorithm \citep{Ma2020UniversalLSM,zhang2022joint} have been developed. As to the frequentist statistical theory for latent space models, \cite{Athreya2018RDPGSurvey} studied spectral embedding estimators under the random dot product graph model. \cite{Ma2020UniversalLSM} considered binary edges with a logistic link function and established an average consistency result for the latent positions. Recently, \cite{Li2023InferenceLSM} proposed a unified and flexible framework for analyzing the theoretical properties of maximum likelihood estimators of latent space models under a general setup. All these works are more focused on the theoretical properties of the estimated latent positions.

This work is inspired by an important property of the network, which is the nodal heterogeneity. It could be observed that a small number of nodes have extremely large degrees, leading to a long-tail pattern of degree distribution \citep{barabasi1999emergence}. Degree heterogeneity can be observed in many real networks, e.g., power grids \citep{albert2002statistical}, social networks \citep{stephen2009explaining}, and others. Various network models have been developed to model degree heterogeneity. The arguably earliest effort is the $p_1$ model of \cite{holland1981exponential}, which utilizes two sets of parameters (i.e., productivity and attractiveness) to control heterogeneity for both out-degree and in-degree. Later, \cite{yan2016asymptotics} try to estimate these heterogeneity parameters in the $p_1$ model using the method of maximum likelihood estimation with guarantees of uniform consistency. Degree heterogeneity is also an important issue for community detection. In this regard, the degree-corrected stochastic block model of \cite{karrer2011stochastic} and its various extensions \citep{sengupta2018block,jin2024mixed} have been extensively studied. Meanwhile, for latent space models, the nodal effect is widely used to account for degree heterogeneity \citep{hoff2003,krivitsky2009representing,chang2019popularity,zhang2022joint,Li2023InferenceLSM,li2025high}.

Although degree heterogeneity has been widely documented in network models, the relationship between degree heterogeneity and nodal features is still not well understood. \cite{Weilan2021influence} propose an adaptive spatial autoregressive model with varying autocorrelation parameters, where nodal influence is parameterized via a regression model on nodal features. \cite{wu2022inward} extend this framework by incorporating additional parameters to quantify their effects on receptive influence, and the same idea has been successfully applied to bipartite networks \citep{wu2024bipartite}. Parameterizing heterogeneity brings two main benefits. First, it improves the interpretability of how nodal features relate to nodal popularity, which is widely observed in real networks. Second, since the number of popularity parameters diverges with network size, but data for each node remains limited, regressing these parameters on covariates effectively reduces the dimensionality of the estimation problem.

To better understand the popularity of the nodes, inspired by the popularity parameter defined in \cite{chang2019popularity}, we develop a novel popularity regression approach by regressing the unobserved popularity parameters on the intended covariates for sparse networks. Unlike \cite{chang2019popularity}, which solely relies on the network structure and provides no theoretical analysis of the estimators, we establish a direct connection between nodal covariates and popularity. This extension enables a rigorous quantification of the effects of covariates on node popularity with interpretable formulations. It also leads to a challenging estimation problem,  sophisticated asymptotic theory involving non-standard $U$-statistics, and surprising theoretical findings, including different convergence rates for the proposed estimators. Our framework allows us to derive explicit expressions for higher-order network properties, which are analytically tractable and therefore computationally easy to implement for downstream tasks. In contrast to the theoretical analysis for LSM, our main focus here is on the estimation of the parameters instead of the latent positions.

Specifically, to avoid the intractable high-dimensional integration required by the maximum likelihood estimator, we develop four estimators based on network features: a pseudo maximum likelihood estimator, an in-degree-based estimator, a reciprocity-based estimator, and a transitivity-based estimator. To study their asymptotic properties, various non-standard $U$-statistics are unfortunately involved. Due to the dependence among network edges, classical projection techniques for $U$-statistics are no longer applicable. To solve this problem, we carefully decompose the network-related summation terms into different martingale difference sequences and develop a new martingale-based technique for network-dependent variables. This ultimately leads to a valid asymptotic theory for these estimators. Surprisingly, these estimators exhibit different asymptotic convergence rates. This is a phenomenon that, to the best of our knowledge, has not been documented in the network analysis literature. We further extend our theory to a general latent position distribution and to estimators built on higher-order network features. We demonstrate the finite-sample performance of all proposed estimators through extensive simulations and a real-data application.

The rest of this article is organized as follows. Section 2 introduces the model structure with a careful discussion of various network properties, such as nodal degree, reciprocity, and transitivity. Section 3 develops a total of four different estimation methods and establishes their asymptotic properties. Section 4 extends the PoRe-LSM for the generalized distribution of latent position and considers estimation based on higher-order statistics. In Section 5, a number of numerical studies are conducted, including both simulation studies and a real data example. Lastly, the article concludes with a brief discussion in Section 6.

\csection{METHODOLOGY}

\csubsection{Model and Notation}

Assume a directed network with $N$ nodes, indexed by $1\leq i\leq N$. Define an adjacency matrix $A=(a_{i_1i_2})\in\mathbb R^{N\times N}$, where $a_{i_1i_2}=1$ if node $i_1$ follows node $i_2$, and $a_{i_1i_2}=0$ otherwise. Moreover, let $a_{ii}=0$ for every $1\leq i\leq N$. In many real-world scenarios, a node's popularity in network is closely related to its attributes, highlighting the importance of incorporating nodal features when modeling network formation \citep{sewell2015latent}. Accordingly, we begin by modeling the relationship between the popularity parameter and the nodal features.
Consider a Twitter-type social network. The popularity of a node (i.e., user) could be closely related to, for example, social status (e.g., whether the user is a celebrity), tenure (e.g., the duration since the registration of the user), and activity (e.g., posting times per week). Define a nodal popularity parameter $\gamma_i\in\mathbb R^+$ for $1\leq i\leq N$. Collect the covariates of node $i$ using a $p$-dimensional vector $X_i=(X_{ij})^\top\in\mathbb R^p$ with $1\leq j\leq p$ and let $\mathbb X=(X_1,\ldots,X_N)^\top\in\mathbb R^{N\times p}$ be the covariate matrix. We then assume that
\begin{align}
\log \gamma_i = X_i^\top\beta+\alpha, \label{eq:pop_reg}
\end{align}
where $\alpha\in\mathbb R$ is the intercept and $\beta=(\beta_j)^\top\in \mathbb R^p$ is the $p$-dimensional coefficient vector associated with the feature vector.

\begin{figure}
    \centering
    \includegraphics[width=0.8\linewidth]{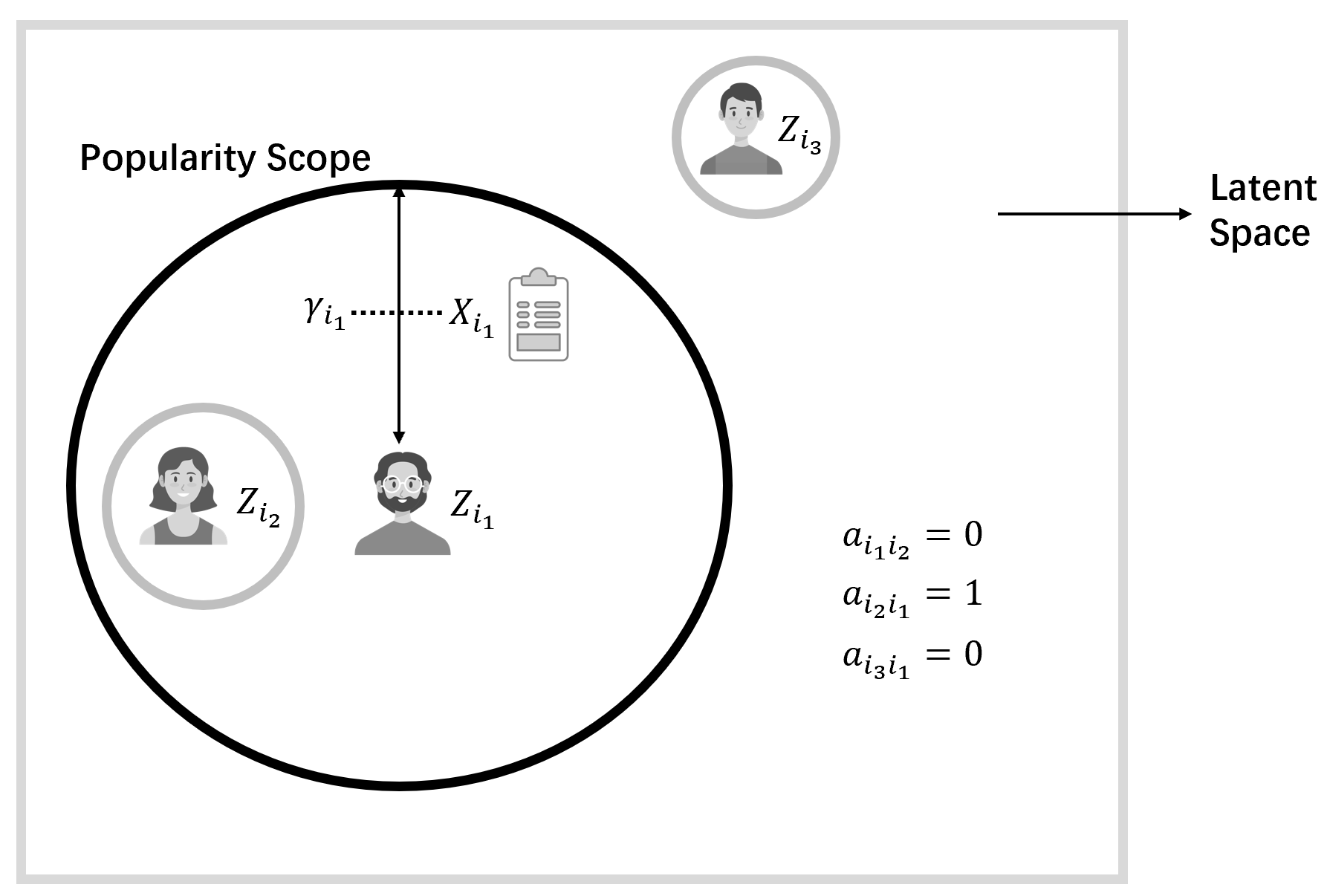}
    \caption{Illustration of PoRe-LSM, where $i_1$, $i_2$, and $i_3$ are three nodes embedded with positions $Z_{i_1}$, $Z_{i_2}$, and $Z_{i_3}$ in latent space respectively. All nodes are associated with a popularity parameter $\gamma$ depending on their nodal feature $X$. Note that $a_{i_1i_2}$, $a_{i_2i_1}$, and $a_{i_3i_1}$ are three binary variables representing the directed edges between $i_1$, $i_2$, and $i_3$, where $a_{i_2i_1}$ is the only directed edge observed in this network.}
    \label{fig:PoRe}
\end{figure}

To model edge generation probability, following \cite{hoff2002latent}, we assume a latent position $Z_i\in\mathbb R^d$ for each node $i$ in the $d$-dimensional Euclidean space. These latent positions are then collected by $\mathbb Z=(Z_1,...,Z_N)\in\mathbb R^{d\times N}$. Once $\mathbb Z$ is given, the edge generation depends on the nodal popularity parameter, and for $1\leq i_1\neq i_2\leq N$,
\begin{align}
P(a_{i_1i_2}=1|\mathbb Z,\mathbb X)=P(a_{i_1i_2}=1|Z_{i_1},Z_{i_2},X_{i_2})=\exp\left(-\frac{\|Z_{i_1}-Z_{i_2}\|^2}{2\gamma_{i_2}^2}\right) \label{eq:latent_model}.
\end{align}
We refer to model \eqref{eq:latent_model}, with the popularity parameter defined in \eqref{eq:pop_reg}, as the Popularity Regression-based Latent Space Model (PoRe-LSM).
 As indicated in model \eqref{eq:latent_model}, the probability that $a_{i_1i_2} = 1$ depends on two key factors: the distance between the latent positions of  $i_1$ and $i_2$, and the popularity of the receiver node $i_2$, which is determined by its nodal features as in model \eqref{eq:pop_reg}. We illustrate the idea of model definition in Figure \ref{fig:PoRe}, where each node is embedded with a latent position $Z$ and associated with a popularity parameter $\gamma$ depending on the nodal feature $X$. Here, the absence of a link $a_{i_3i_1} = 0$ is primarily due to the limited popularity scope of node $i_1$, which is insufficient to reach the distance between $Z_{i_1}$ and $Z_{i_3}$. Moreover, we observe that $a_{i_1i_2} = 0$ while $a_{i_2i_1} = 1$, indicating that the popularity scope of node $i_1$ extends to the latent position $Z_{i_2}$, whereas the scope of node $i_2$ does not cover $Z_{i_1}$. Furthermore, instead of employing a hard threshold based on $\|Z_{i_1} - Z_{i_2}\|/\gamma_{i_2}$, we adopt a probabilistic model in \eqref{eq:latent_model} serving as a soft threshold for determining the presence of an edge $a_{i_1i_2} = 1$. For simplicity, we assume that the $Z_i$s are independently drawn from a standard normal distribution. All the results can be extended to the general $d$-dimensional setting.

As pointed out by many researchers, a large-scale network should have a sparse structure \citep{watts1998collective}. To this end, assuming $(\alpha,\beta^\top)^\top$ and thus $\gamma_i$ as fixed constants as $N\to\infty$ for every $1\leq i\leq N$ might be inappropriate, since this leads to a non-sparse network structure \citep{schweinberger2015local}. To generate a sparse network structure, it is essential to require that $\gamma_i \to 0$ as $N\to\infty$ at an appropriate rate. Unfortunately, this desired theoretical property can hardly be achieved by enforcing $\|\beta\| \to \infty$ as $N\to\infty$. This is because the sign of $X_i$ could be positive or negative. Therefore, it seems more appropriate to enforce an assumption on the intercept $\alpha$ and replace it with $\alpha_N$, where the subscript is to emphasize the fact that this is a parameter depending on the network size $N$. For notation simplicity, we let $\theta = (\alpha_N,\beta^\top)^\top\in\mathbb R^{p+1}$. Next, following \cite{wang2020logistic}, we assume $\alpha_N\to-\infty$ as $N\to\infty$ at an appropriate speed. As a result, it is necessary to carefully study what kind of divergence rate can be viewed as appropriate.

\csubsection{Nodal Degree in PoRe-LSM}

In this subsection, we focus on the properties of nodal degree and study the divergence rate of $\alpha_N$. Define the out-degree and in-degree of node $i$ as $d^\text{out}_i=\sum_{i^\prime=1}^N a_{ii^\prime}$ and $d_i^\mathrm{in}=\sum_{i^\prime=1}^Na_{i^\prime i}$. We begin with a discussion of the asymptotic behavior of $E(d^\text{out}_i)$.
 For many networks with an extremely sparse structure, we should expect $E(d^\text{out}_i)\leq C_{\max}$ as $N\to\infty$ for some fixed positive constant $C_{\max}$. If the network structure is slightly denser but still sparse, we might also expect $E(d^\text{out}_i)\to\infty$ as $N\to\infty$. This is mainly because a diverging network size $N$ does give every node an increasingly large number of candidate nodes to follow. Furthermore, since each node (e.g., a social media user) only has a limited attention budget, the divergence rate of $E(d^\text{out}_i)$ should be sufficiently slow in the sense $E(a_{i_1i_2})\to0$ for any $i_1\neq i_2$ as $N\to\infty$.

 Theoretically, this amounts to assuming a slowly diverging upper bound for the expected out-degree as $N^\delta C_{\min} \leq E(d^{\text{out}}_i)\leq N^{\delta}C_{\max}$ as $N\to\infty$, where $C_{\min}\leq C_{\max}$ are two positive constants and $\delta$ is a sufficiently small and fixed nonnegative constant. By the definition of PoRe-LSM in \eqref{eq:pop_reg} and \eqref{eq:latent_model}, it follows then
\begin{align*}
E\Big(d_i^{\text{out}}\Big) &=\sum_{i^\prime\neq i}^N E\Big\{E\big(a_{ii^\prime}\big|\mathbb X\big)\Big\}= \frac{N\exp(\alpha_N)}{\sqrt{2}}E\Big\{\exp\left(X_i^\top\beta\right)\Big\}\Big\{1+o(1)\Big\},
\end{align*}
where the detailed derivation of $E(d_i^{\text{out}})$ is referred to Appendix S1.1. Then, write $E\{\exp(X_i^\top\beta)\}=C_\beta$. Therefore, to ensure $N^\delta C_{\min}\leq E(d_i^{\text{out}})\leq  N^{\delta}C_{\max}$, we should have $\sqrt{2}C_{\min}/C_\beta\leq N^{1-\delta}\exp(\alpha_N)\leq\sqrt{2}C_{\max}/C_\beta$ as $N\to\infty$. To meet this theoretical requirement, it is natural and convenient to assume that $N^{1-\delta}\exp(\alpha_N)\to C_\alpha$ for some positive constant $C_\alpha\in [\sqrt{2}C_{\min}/C_\beta,\sqrt{2}C_{\max}/C_\beta]$. Consequently, we should assume that $\alpha_N = \log C_\alpha-(1-\delta)\log N+o(1)$ with $0\leq\delta<1$. Note that as $\delta$ increases, the expected out-degree $E(d^{\text{out}}_i)$ increases and thus the whole network becomes denser. Therefore, we refer to $\delta$ as the density level of the network for the rest of this paper.

\noindent
{\bf Remark 1.} For $d_i^{\mathrm{in}}$, we have $E(d_i^{\mathrm{in}})=\sum_{i^\prime\neq i}E(a_{i^\prime i})=(N-1)E\{\gamma_i(2+\gamma_i^2)^{-1/2}\}=2^{-1/2}N\exp(\alpha_N)E\{\exp(X_i^\top\beta)\}\{1+o(1)\}$. Thus, $E(d_i^{\text{out}})$ and $E(d_i^{\mathrm{in}})$ have the same order as the network size $N\to\infty$. Therefore, when the network is extremely sparse, we should also expect $E(d_i^{\mathrm{in}})\leq C_{\max}$. For a network that is slightly denser but overall still sparse, we should expect that $E(d_i^\mathrm{in})\to\infty$ at the same sufficiently slow rate as $E(d_i^\text{out})$ when $N\to\infty$. This implies that each node has more chances to be followed by others, but the number of followers cannot be too large on average.

It is widely acknowledged that nodal in-degree and out-degree represent popularity and activity, respectively \citep{wasserman1994social}. We should expect that nodal in-degree is more closely related to the popularity parameter $\gamma_i$, as compared with the out-degree. To illustrate this idea, we study the conditional expectation of nodal degree (i.e., both $d_i^{\mathrm{in}}$ and $d_i^{\text{out}}$) given nodal feature $\mathbb X$. By PoRe-LSM defined in \eqref{eq:pop_reg} and \eqref{eq:latent_model}, we have,
\begin{align}
&E(d_i^{\mathrm{in}}|\mathbb X)=\sum\limits_{i^\prime\neq i}^N E(a_{i^\prime i}|\mathbb X)=\frac{N\exp(X_i^\top\beta+\alpha_N)}{\sqrt{2}}\bigg\{1+o_p(1)\bigg\},\label{eq:in_degree}\\
&E(d_i^{\text{out}}|\mathbb X)=\sum\limits_{i^\prime\neq i}^N E(a_{ii^\prime }|\mathbb X)=\sum_{i^\prime\neq i}^N\frac{\exp(X_{i^\prime}^\top\beta+\alpha_N)}{\sqrt{2}}\bigg\{1+o_p(1)\bigg\},\label{eq:out_degree}
\end{align}
where the detailed derivation is in Appendix S1.1. The conditional expectation of out-degree of node $i$ is related to all other nodes through $X_{i'}^\top\beta$ with $i'\neq i$. On the other hand, the conditional expectation of in-degree of node $i$ is directly related to $X_i$ through $X_i^\top\beta$. It indicates that the conditional expectation of nodal in-degree can be utilized for parameter estimation. In the next section, we develop an in-degree-based estimator for the regression coefficient $\beta$ and the intercept parameter $\alpha_N$ in PoRe-LSM.

\csubsection{Reciprocity and Transitivity in PoRe-LSM}

Nodal degree is a first-order type of network structure. In this subsection, we consider two higher-order network structures, i.e., reciprocity and transitivity. Reciprocity refers to the phenomenon of mutual follow-up among nodes (i.e., $a_{i_1i_2}=a_{i_2i_1} = 1$). It is an interesting property widely observed in large-scale and sparse networks \citep{holland1981exponential,strauss1990pseudolikelihood}. Specifically, given $a_{i_1i_2}=1$, the probability of $a_{i_2i_1}=1$ should be considerably larger than the unconditional probability of $a_{i_2i_1}=1$. Mathematically, this amounts to investigating the conditional probability $P(a_{i_2i_1}=1|a_{i_1i_2}=1,\mathbb X)=P(a_{i_2i_1}a_{i_1i_2}=1|\mathbb X)/P(a_{i_1i_2}=1|\mathbb X)$. Recall that $P(a_{i_1i_2}=1|\mathbb X)=\gamma_{i_2}(\gamma_{i_2}^2+2)^{-1/2}$. We next study the value of $P(a_{i_2i_1}a_{i_1i_2}=1|\mathbb X)$. By \eqref{eq:pop_reg} and \eqref{eq:latent_model}, we have
$P(a_{i_2i_1}a_{i_1i_2}=1|\mathbb X)=\gamma_{i_1}\gamma_{i_2}(\gamma_{i_1}^2\gamma_{i_2}^2+2\gamma_{i_1}^2+2\gamma_{i_2}^2)^{-1/2}$, where the detailed derivation is referred to Appendix S1.1. By straightforward calculations, it can be verified that $P(a_{i_2i_1}=1|a_{i_1i_2}=1,\mathbb X)=$
\begin{align}
&\bigg\{\frac{\gamma_{i_1}\gamma_{i_2}}{(\gamma_{i_1}^2\gamma_{i_2}^2+2\gamma_{i_1}^2+2\gamma_{i_2}^2)^{1/2}}\bigg\}\bigg\{\frac{\gamma_{i_2}}{(\gamma_{i_2}^2+2)^{1/2}}\bigg\}^{-1}=\sqrt{\frac{\exp(2\Delta_{i_1i_2}^\top\beta)}{\exp(2\Delta_{i_1i_2}^\top\beta)+1}}\Bigg\{1+o_p(1)\Bigg\},\label{eq:cond_recip}
\end{align}
where $\Delta_{i_1i_2}=X_{i_1}-X_{i_2}\in\mathbb R^p$ and the equation is due to the fact that $\gamma_i\rightarrow_p 0$ as $N\rightarrow \infty$. It can be seen that this conditional probability is bounded above 0 by a fixed constant depending on $\mathbb X$ as $N\to\infty$. This interesting finding suggests that although the probability of $a_{i_1i_2}=1$ tends towards 0 in a large-scale network, once it is given that $i_2$ follows $i_1$ (i.e., $a_{i_2i_1}=1$), the conditional probability of $a_{i_1i_2}=1$ is no longer trivial. In the next section, we utilize this useful property and develop a reciprocity-based estimator for PoRe-LSM.

Transitivity, another important high-order network structure, involves three distinct nodes and their relationships. Specifically, if node $i_1$ follows node $i_2$ and node $i_2$ follows node $i_3$, there will be a higher probability that node $i_1$ also follows node $i_3$. The phenomenon of transitivity is extensively observed in large-scale networks, reflecting that relationships tend to cluster \citep{handcock2007model}. Mathematically, transitivity means that $a_{i_1i_2}a_{i_2i_3}a_{i_1i_3}=1$. Similar to $P(a_{i_2i_1}a_{i_1i_2}=1|\mathbb X)$, we can spell out the conditional probability $P(a_{i_1i_2}a_{i_2i_3}a_{i_1i_3}=1|\mathbb X)=\gamma_{i_2}\gamma_{i_3}^2(3\gamma_{i_2}^2+6\gamma_{i_3}^2+2\gamma_{i_3}^4+4\gamma_{i_2}^2\gamma_{i_3}^2+\gamma_{i_2}^2\gamma_{i_3}^4)^{-1/2}$ and $P(a_{i_1i_2}a_{i_2i_3}=1|\mathbb X)=\gamma_{i_2}\gamma_{i_3}(3+2\gamma_{i_2}^2+2\gamma_{i_3}^2+\gamma_{i_2}^2\gamma_{i_3}^2)^{-1/2}$. Note that $P(a_{i_1i_3}=1|a_{i_1i_2}a_{i_2i_3}=1,\mathbb X)=P(a_{i_1i_2}a_{i_2i_3}a_{i_1i_3}=1|\mathbb X)/P(a_{i_1i_2}a_{i_2i_3}=1|\mathbb X)$, we have $P(a_{i_1i_3}=1|a_{i_1i_2}a_{i_2i_3}=1,\mathbb X)=$
\begin{equation}
\frac{(3\gamma_{i_3}^2+2\gamma_{i_2}^2\gamma_{i_3}^2+2\gamma_{i_3}^4+\gamma_{i_2}^2\gamma_{i_3}^4)^{1/2}}{(3\gamma_{i_2}^2+6\gamma_{i_3}^2+2\gamma_{i_3}^4+4\gamma_{i_2}^2\gamma_{i_3}^2+\gamma_{i_2}^2\gamma_{i_3}^4)^{1/2}}=\sqrt{\frac{\exp(2\Delta_{i_3i_2}^\top\beta)}{2\exp(2\Delta_{i_3i_2}^\top\beta)+1}}\Bigg\{1+o_p(1)\Bigg\}\label{eq:cond_trans}.
\end{equation}
It can be seen that this conditional probability is also bounded above 0 by a constant depending on $\mathbb X$ as $N\to\infty$. This indicates that given $a_{i_1i_2}a_{i_2i_3}=1$, the conditional probability of observing the transitivity structure is not trivial. In the next section, we also develop a transitivity-based estimator based on the transitivity property.

 \csection{PARAMETER ESTIMATION}

In this section, we consider parameter estimation of the interested regression coefficient $\beta$ and also the nuisance parameter $\alpha_N$ in PoRe-LSM. We propose two first-order-based estimators and two higher-order-based estimators.

\csubsection{Pseudo Maximum Likelihood Estimator}

We start with a pseudo maximum likelihood estimator (PMLE). Specifically, by \eqref{eq:latent_model}, we show that the conditional probability of $a_{i'i}=1$ given $\mathbb X$ is
\beqr
P(a_{i'i}=1|\mathbb X)=\gamma_i(\gamma_i^2+2)^{-1/2}=\gamma_i\{1+o_p(1)\}/\sqrt{2}.\label{eq:conprob}\eeqr Therefore, if we
treat $a_{i_2i_1}$s as if they were independent binary random variables with response probability $\gamma_i/\sqrt{2}$, a pseudo log-likelihood function can be constructed as $\widetilde{\mathcal L}_{\mathrm{pmle}}(\theta)=\sum_{i_1\neq i_2}\{a_{i_2i_1}\log (\gamma_{i_1}/\sqrt{2}) + (1-a_{i_2i_1})\log (1-\gamma_{i_1}/\sqrt{2})\}$. Recall that $\widetilde X_i=(1,X_i^\top)^\top$, $\theta = (\alpha_N,\beta^\top)^\top$. By Taylor's expansion,  we have
\begin{align*}
\widetilde{\mathcal L}_{\mathrm{pmle}}(\theta)=\sum_{i=1}^N \left\{d_{i}^\mathrm{in}\widetilde X_i^\top\theta-\big(N-1-d_{i}^\mathrm{in}\big)\exp\big(\widetilde X_i^\top\theta\big)/\sqrt{2}\right\}\Big\{1+o_p(1)\Big\},
\end{align*}
where the equality is due to the fact that $\gamma_i=o_p(1)$ and holds up to some constant irrelevant to $\theta$. We refer the detailed derivation to Appendix S1.1. Therefore, this leads to the following loss function
\begin{align}
\mathcal L_{\mathrm{pmle}}(\theta)=-\sum_{i=1}^N \left\{d_{i}^\mathrm{in}\widetilde X_i^\top\theta-\Big(N-1-d_{i}^\mathrm{in}\Big)\exp\big(\widetilde X_i^\top\theta\big)/\sqrt{2}\right\}.\label{eq:plmelik}
\end{align}
Then a natural and tractable estimator for $
\theta$ can be defined as $\widehat \theta_{\mathrm{pmle}}=\arg\min_{\theta}\mathcal L_{\mathrm{pmle}}(\theta)$. To study the asymptotic property of $\widehat \theta_{\mathrm{pmle}}$, the following conditions are needed.
\begin{itemize}
\item[(C1)] (Nodal Features and Latent Positions) Assume $X_i$s are independent and identically distributed such that $E\{\exp(X_i^\top\beta)\}<\infty$ for all fixed $\beta\in\mathbb R^p$. Furthermore, assume that the latent position $Z_i$s are independent standard normal random variables and $a_{i_1i_2}$s are conditionally independent given $\mathbb X$ and $\mathbb Z$.
\item[(C2)] (Bound of Density Level) Assume $\alpha_N = \log C_\alpha-(1-\delta)\log N + o(1)$, where network density level $\delta \in[0,1/2)$ and $C_\alpha$ is some fixed positive constant.
\end{itemize}
Condition (C1) is basically a moment condition for $X_i$, which can be easily satisfied if $X_i$s are of a sub-Gaussian distribution. It is remarkable that by Condition (C1), we have $\gamma_{i_1}/\gamma_{i_2}=\exp (X_{i_1}^\top\beta-X_{i_2}^\top\beta) =O_p(1)$ for any $i_1 \neq i_2$.
However, this never implies that $\gamma_{\max}/\gamma_{\min} = O_p(1)$,
where $\gamma_{\max} = \max_{1 \leq i \leq N}\{\gamma_i\} = \exp\{\max_{1 \leq i \leq N}(X_i^\top \beta)\}$
and $\gamma_{\min} = \exp\{\min_{1 \leq i \leq N}(X_i^\top \beta)\}$.
In fact, we have
$\gamma_{\max}/\gamma_{\min}
= \exp\{\max_{1 \leq i \leq N}(X_i^\top \beta) - \min_{1 \leq i \leq N}(X_i^\top \beta)\}.$
This is a random quantity, which is allowed to diverge to infinity in probability easily,
as long as $\beta \neq 0$ and $X_i$ has some unbounded support. Condition (C2) implies that the expected in-degree $E(d_i^{\mathrm{in}})$ and out-degree $E(d_i^{\text{out}})$ diverge as $N\to\infty$ but at a relatively slow rate. Theoretically, the upper bound of $\delta<1/2$ controls the bias introduced by the difference between the true log-likelihood function and the proposed pseudo one. Under these conditions, we have the following theorem.

\begin{theorem}[Asymptotic Properties of PMLE]
    Assume that conditions (C1) and (C2) hold. We then have: (1) $N^{1/2}(\widehat\theta_{\mathrm{pmle}}-\theta)\xrightarrow[]{d} N(0,\mathbf H_{\mathrm{pmle}}^{-1}\mathbf M_{\mathrm{pmle},1}\mathbf H_{\mathrm{pmle}}^{-1})$ if $\delta>0$ and (2) $N^{1/2}(\widehat\theta_{\mathrm{pmle}}-\theta)\xrightarrow[]{d} N(0,\mathbf H_{\mathrm{pmle}}^{-1}\mathbf M_{\mathrm{pmle},0}\mathbf H_{\mathrm{pmle}}^{-1})$ if $\delta=0$, where $\mathbf H_{\mathrm{pmle}}=E\{\exp(X_i^\top\beta)\widetilde X_i \widetilde X_i^\top\}/\sqrt{2}\in\mathbb R^{(p+1)\times (p+1)}$. Moreover, $\mathbf M_{\mathrm{pmle},0}\in\mathbb R^{(p+1)\times (p+1)}$ and $\mathbf M_{\mathrm{pmle},1}\in\mathbb R^{(p+1)\times (p+1)}$ are two positive definite matrices defined as
\begin{align*}
&\mathbf M_{\mathrm{pmle},1}=\left(\sqrt{3}-\frac{3}{2}\right)E\left(e^{X_i^\top\beta}\widetilde X_i \right)E\left(e^{X_i^\top\beta}\widetilde X_i^\top\right)+\left(\frac{1}{\sqrt{3}}-\frac{1}{2}\right)E\left(e^{2X_i^\top\beta}\widetilde X_i\widetilde X_i^\top\right),\\
&\mathbf M_{\mathrm{pmle},0}=\mathbf M_{\mathrm{pmle},1}+(\sqrt{2}C_\alpha)^{-1}E\big(e^{X_i^\top\beta}\widetilde X_i\widetilde X_i^\top\big)+(\sqrt{2}C_\alpha)^{-1}E\left(\frac{e^{X_{i_1}^\top\beta+X_{i_2}^\top\beta}\widetilde X_{i_1}\widetilde X_{i_2}^\top}{\sqrt{e^{2X_{i_1}^\top\beta}+e^{2X_{i_2}^\top\beta}}}\right),
\end{align*}
where $i_1\neq i_2$ are two different indices.
\end{theorem}
\noindent
The proof of Theorem 1 is given in Appendix S2.1. Our parameter estimation method is carried out marginally by integrating over the latent positions $\{Z_i\}_{i=1}^N$. Accordingly, our method cannot provide any ``explainability'' in terms of the latent position $\{Z_i\}_{i=1}^N$. For valid statistical inference for $\widehat\theta_{\mathrm{pmle}}$, $\mathbf M_{\mathrm{pmle},0}$, $\mathbf M_{\mathrm{pmle},1}$ and $\mathbf H_{\mathrm{pmle}}$ in Theorem 1 need to be consistently estimated. The analytical details of these quantities are given in Appendix S3.1. This enables straightforward plug-in type consistent estimators. Moreover, we show in Appendix S3.1 that $\mathbf M_{\mathrm{pmle},0}$ is a positive definite matrix and $\mathbf M_{\mathrm{pmle},0}-\mathbf M_{\mathrm{pmle},1}$ is a semi-positive definite matrix. This implies that the estimator obtained with $\delta>0$ is statistically more efficient than the one with $\delta=0$.

When $\delta=0$, we should expect that the total number of edges is $E(\sum_{i_1\neq i_2} a_{i_1i_2})=N(N-1)E\{\gamma_{i}(\gamma_{i}^2+2)^{-1/2}\}=CN\{1+o(1)\}$ for some constant $C$. Therefore, it is reasonable to have $\widehat\beta_{\mathrm{pmle}}$ be $N^{1/2}$-consistent. However, with $\delta>0$, we should have $E(\sum_{i_1\neq i_2} a_{i_1i_2})=CN^{1+\delta}\{1+o(1)\}$. Therefore, we should expect the convergence rate of a reasonable estimator to be $N^{(1+\delta)/2}$, which is a rate faster than $N^{1/2}$. Unfortunately, by Theorem 1, we find that $\widehat \beta_{\mathrm{pmle}}$ remains $N^{1/2}$-consistent, which is not optimal. Moreover, our technical proof given in Appendix S2.1 reveals that the asymptotic bias of $N^{1/2}(\widehat \beta_{\mathrm{pmle}}-\beta)$ is of order $N^{\delta-1/2}$, which converges towards 0 at a painfully slow rate. The consequence is that the empirical coverage rate of the estimated confidence interval according to Theorem 1 converges to nominal level at an extremely slow rate when $\delta>0$. This will be demonstrated in the simulation study. In Appendix S2.1, we further show that the PMLE $\widehat\theta_{\mathrm{pmle}}$ is in fact consistent for $\theta$ but not necessarily asymptotically normal for any $\delta\in[0,1)$. This is mainly because the asymptotic bias might dominate its asymptotic standard deviation if $\delta\in[1/2,1)$.

\csubsection{An In-Degree-Based Estimator}

The unsatisfactory performance of PMLE when $\delta>0$ inspires us to explore an alternative first-order estimator that potentially provides improved finite sample performance. Since this is an estimator constructed based on nodal in-degree, we refer to it as an in-degree-based estimator (IN). Specifically, we consider the regression relationship between $\mathbb X$ and the nodal in-degree. This is because nodal in-degree is more related to the popularity parameter $\gamma_i$ defined in PoRe-LSM \eqref{eq:pop_reg} and \eqref{eq:latent_model}, as compared with nodal out-degree. Analytically, from equation \eqref{eq:out_degree}, it can be seen that the value of $\gamma_i$ is directly related to $X_i$ through $X_i^\top\beta$. In contrast, that of the out-degree is independent of $X_i$ but related to $X_{i'}$ with $i'\not = i$. Therefore, nodal out-degree is inappropriate for parameter estimation. Subsequently, we develop a least squares type of loss function of $\theta$ based on nodal in-degree as
\begin{align}
\mathcal L_{\mathrm{in}}(\theta) =\sum_{i=1}^N\left\{d_i^{\mathrm{in}}-\frac{N\exp(\widetilde X_i^\top\theta)}{\sqrt{2}}\right\}^2=\sum_{i=1}^N\ell_{\mathrm{in}}^i(\theta),\label{eq:inlik}
\end{align}
where $\widetilde X_i=(1,X_i^\top)^\top\in\mathbb R^{p+1}$ and $\ell_{\mathrm{in}}^i(\theta) = \{d_i^{\mathrm{in}}-N\exp(\widetilde X_i^\top\theta)/\sqrt{2}\}^2$ for $1\leq i\leq N$. The loss function in \eqref{eq:inlik} leads to an estimator $\widehat \theta_{\mathrm{in}}=\arg\min_\theta\mathcal L_{\mathrm{in}}(\theta)$ and the asymptotic property of $\widehat\theta_{\mathrm{in}}$ is given by the following theorem. It is remarkable that, in the construction of the loss functions of both PMLE and IN estimators, the approximation $\gamma_i(\gamma_i^2+2)^{-1/2}=\gamma_i\{1+o_p(1)\}/\sqrt{2}$ has been used for simplicity; see \eqref{eq:conprob}. In fact, we can also use the accurate formula $\gamma_i(\gamma_i^2+2)^{-1/2}$ directly for both PMLE and IN. The resulting estimators share the same asymptotic distribution as their approximated counterparts. The theoretical and numerical verification details for IN are given in Appendix S3.3 and Appendix S4.3, respectively. The verification details for PMLE are very similar and therefore are not presented to save space.

\begin{theorem}[Asymptotic Properties of IN]
   Assume conditions (C1) and (C2) hold. We then have: (1) $N^{1/2}(\widehat\theta_{\mathrm{in}}-\theta)\xrightarrow[]{d} N(0,\mathbf H_{\mathrm{in}}^{-1}\mathbf M_{\mathrm{in},1}\mathbf H_{\mathrm{in}}^{-1})$ if $\delta>0$ and (2) $N^{1/2}(\widehat\theta_{\mathrm{in}}-\theta)\xrightarrow[]{d} N(0,\mathbf H_{\mathrm{in}}^{-1}\mathbf M_{\mathrm{in},0}\mathbf H_{\mathrm{in}}^{-1})$ if $\delta=0$, where $\mathbf H_{\mathrm{in}}=E\{\exp(2X_i^\top\beta)\widetilde X_i \widetilde X_i^\top\}$, $\mathbf M_{\mathrm{in},0}\in\mathbb R^{(p+1)\times (p+1)}$ and $\mathbf M_{\mathrm{in},1}\in\mathbb R^{(p+1)\times (p+1)}$ are two positive definite matrices defined as
\begin{align*}
&\mathbf M_{\mathrm{in},1}=\Big(2\sqrt{3}-3\Big)E\left(e^{2X_i^\top\beta}\widetilde X_i \right)E\left(e^{2X_i^\top\beta}\widetilde X_i^\top\right)+\Big(2/\sqrt{3}-1\Big)E\left(e^{4X_i^\top\beta}\widetilde X_i\widetilde X_i^\top\right),\\
&\mathbf M_{\mathrm{in},0}=\mathbf M_{\mathrm{in},1}+\sqrt{2}C_\alpha^{-1}E\left(e^{3X_i^\top\beta}\widetilde X_i\widetilde X_i^\top\right)+\sqrt{2}C_\alpha^{-1}E\left(\frac{e^{2X_{i_1}^\top\beta+2X_{i_2}^\top\beta}\widetilde X_{i_1}\widetilde X_{i_2}^\top}{\sqrt{e^{2X_{i_1}^\top\beta}+e^{2X_{i_2}^\top\beta}}}\right),
\end{align*}
where $i_1\neq i_2$ are two different indices.
\end{theorem}
\noindent The proof of Theorem 2 is given in Appendix S2.2. Similar to Theorem 1, it can be verified in Appendix S3.1 that $\mathbf M_{\mathrm{in},0}-\mathbf M_{\mathrm{in},1}$ is a semi-positive definite matrix and the estimator obtained with $\delta>0$ is statistically more efficient than the one with $\delta=0$. Unfortunately, by Theorem 2, we find that $\widehat\theta_{\mathrm{in}}$ remains to be $N^{1/2}$-consistent regardless of the value of $\delta$. Compared with $\widehat\theta_{\mathrm{pmle}}$, as revealed in Appendix S2.2, we find that the asymptotic bias of $N^{1/2}(\widehat\theta_{\mathrm{in}}-\theta)$ is of the order $N^{2\delta-3/2}$, which converges towards 0 at a faster rate than $N^{\delta-1/2}$ of $N^{1/2}(\widehat\theta_{\mathrm{pmle}}-\theta)$. Therefore, the empirical coverage rate of $\widehat\theta_{\mathrm{in}}$ is better than that of $\widehat\theta_{\mathrm{pmle}}$ when $\delta$ is relatively large, which will be demonstrated in the subsequent simulations. For valid statistical inference for $\widehat\theta_{\mathrm{in}}$, $\mathbf M_{\mathrm{in},0}$, $\mathbf M_{\mathrm{in},1}$ and $\mathbf H_{\mathrm{in}}$ in Theorem 2 need to be consistently estimated. Details are given in Appendix S3.1.

\noindent {\bf Remark 2. } The results of both Theorem 1 and Theorem 2 show that $\widehat\beta-\beta$ and $\widehat\alpha-\alpha_N$ converge at the same rate. As we pointed out in Appendix S2.1 and Appendix S2.2, diverging $\alpha_N$ changes the convergence rate of both $\dot{\mathcal L}(\theta)$ and $\ddot{\mathcal L}(\theta)$ through $\delta$ as $N\to\infty$ but cancel each other. Moreover, all the elements in $\dot{\mathcal L}(\theta)$ or $\ddot{\mathcal L}(\theta)$ have the same stochastic order. This leads to a unified convergence rate of $\widehat\beta-\beta$ and $\widehat\alpha-\alpha_N$. Similar phenomena are observed in a rare event setting, where fixed regression coefficients and a diverging intercept have the same convergence rate \citep{wang2020logistic}.

\csubsection{A Reciprocity-Based Estimator}

As can be seen, the convergence rate of $\widehat\beta_{\mathrm{in}}$ remains unsatisfactory with $N^{1/2}$ even when $\delta>0$. One possible reason for such an unsatisfactory convergence rate of both $\widehat\beta_{\mathrm{pmle}}$ and $\widehat\beta_{\mathrm{in}}$ could be the usage of first-order information (i.e., $a_{i_1i_2}$s) only. Here, the first-order information refers to the fact that the loss functions $\mathcal L_{\mathrm{pmle}}(\theta)$ and $\mathcal L_{\mathrm{in}}(\theta)$ are linear functions in $a_{i_1i_2}$s. On the other hand, it has been widely demonstrated in the past literature that large-scale networks often exhibit higher-order structures, which contain rich information for regression analysis \citep{holland1981exponential,frank1986markov,hoff2002latent}. Therefore, we are motivated to develop estimators for PoRe-LSM that utilize higher-order information to achieve an improved convergence rate. In this case, we should construct loss functions, which are higher order (not linear) functions in $a_{i_1i_2}$s.

To this end, in this subsection, we study a reciprocity-based estimator (RE). Recall that in equation \eqref{eq:cond_recip}, we obtain the conditional probability $P(a_{i_2i_1}=1|a_{i_1i_2}=1,\mathbb X)$ for any $i_1\neq i_2$. Practically, this suggests that the conditional probability that node $i_2$ follows node $i_1$ given node $i_2$ followed by node $i_1$ is approximately equal to $\exp(\Delta_{i_1i_2}^\top\beta)/\{\exp(2\Delta_{i_1i_2}^\top\beta)+1\}^{1/2}$ with $\Delta_{i_1i_2}=X_{i_1}-X_{i_2}$. This approximate conditional probability is free of the nuisance parameter $\alpha_N$ and therefore does not shrink toward 0 as $N\to\infty$. Specifically, let $\pi^{i_1i_2}_{\mathrm{re}}$ be the approximated conditional probability of $a_{i_2i_1}=1$ given $a_{i_1i_2}=1$ and $\mathbb X$,
\begin{align}
\pi^{i_1i_2}_{\mathrm{re}}=\frac{\gamma_{i_1}}{(\gamma_{i_1}^2+\gamma_{i_2}^2)^{1/2}}=\frac{\exp(\Delta_{i_1i_2}^\top\beta)}{\{\exp(2\Delta_{i_1i_2}^\top\beta)+1\}^{1/2}}.\label{eq:recip_prob}
\end{align}
Thus, we focus on those $(i_1,i_2)$ pairs with $a_{i_1i_2}=1$ and then treat $a_{i_2i_1}$ as if they were independent binary random variables with response probability $\pi^{i_1i_2}_{\mathrm{re}}$. This leads to the following loss function
\begin{align}
\mathcal L_{\mathrm{re}}(\beta) = -\sum_{i_1\neq i_2}a_{i_1i_2}\Big\{a_{i_2i_1}\log \pi^{i_1i_2}_{\mathrm{re}}+\big(1-a_{i_2i_1}\big)\log\big(1-\pi^{i_1i_2}_{\mathrm{re}}\big)\Big\},\label{eq:relik}
\end{align}
where the detailed derivation is given in Appendix S1.1. As one can see, this loss function \eqref{eq:relik} is a quadratic function in $a_{i_1i_2}$s. Therefore, the second-order information of network can be utilized. Accordingly, a natural and tractable estimator for $\beta$ can be defined as $\widehat\beta_{\mathrm{re}}=\arg\min_\beta\mathcal L_{\mathrm{re}}(\beta)$. The asymptotic property of $\hat\beta_{\mathrm{re}}$ is given below.
\begin{theorem}[Asymptotic Properties of RE]
       Assume conditions (C1) and (C2) hold. We then have $N^{(1+\delta)/2}(\widehat\beta_{\mathrm{re}}-\beta)\xrightarrow[]{d} N(0,C_\alpha^{-1}\mathbf H^{-1}_{\mathrm{re}}\mathbf M_{\mathrm{re}}\mathbf H^{-1}_{\mathrm{re}})$, where
       $\mathbf M_{\mathrm{re}}\in\mathbb R^{p\times p}$ and
$\mathbf H_{\mathrm{re}}\in\mathbb R^{p\times p}$ are two positive definite matrices defined as follows,
\begin{align*}
\mathbf M_{\mathrm{re}}=&E\left[\exp\Big(X_{i_1}^\top\beta\Big)\Big(1-{\pi_{\mathrm{re}}^{i_1i_2}}^2\Big)\Big(\pi_{\mathrm{re}}^{i_1i_2}+{\pi_{\mathrm{re}}^{i_2i_1}}^2\Big)\Big\{2\exp\big(2\Delta_{i_1i_2}^\top\beta\big)+2\Big\}^{-1/2}\Delta_{i_1i_2}\Delta_{i_1i_2}^\top\right],\\
\mathbf H_{\mathrm{re}}&= E\left[\pi_{\mathrm{re}}^{i_1i_2}\Big(\pi_{\mathrm{re}}^{i_1i_2}+1\Big)\exp\Big(X_{i_2}^\top\beta\Big)\Big\{\sqrt{2}+\sqrt{2}\exp\big(2\Delta_{i_1i_2}^\top\beta\big)\Big\}^{-1}\Delta_{i_1i_2}\Delta_{i_1i_2}^\top\right],
\end{align*}
where $i_1\neq i_2$ are two different indices.
\end{theorem}
\noindent The proof of Theorem 3 is given in Appendix S2.3. By Theorem 3, we find that $\widehat\beta_{\mathrm{re}}$ is $N^{(1+\delta)/2}$-consistent and asymptotically normal with analytically tractable asymptotic covariance. This is indeed a faster convergence rate compared to those of $\widehat\beta_{\mathrm{pmle}}$ and $\widehat\beta_{\mathrm{in}}$ when $\delta>0$. For valid statistical inference for $\widehat\beta_{\mathrm{re}}$, $\mathbf M_{\mathrm{re}}$ and $\mathbf H_{\mathrm{re}}$ in Theorem 3 need to be consistently estimated. The technical details are given in Appendix S3.1.

\csubsection{A Transitivity-Based Estimator}

The convergence rate of $\widehat\beta_{\mathrm{re}}$ is indeed promising, which means the higher-order network structures can be extremely useful for PoRe-LSM. Note that $\widehat\beta_{\mathrm{re}}$ is a statistic based on second-order network information (i.e., reciprocity), incorporating quantities such as $a_{i_1i_2}a_{i_2i_1}$. It is then natural to question: can the efficiency of the estimator be further improved by incorporating the third-order network structures? This leads to the following transitivity-based estimator (TR) for PoRe-LSM. By equation \eqref{eq:cond_trans}, we obtain the conditional probability $P(a_{i_1i_3}=1|a_{i_1i_2}a_{i_2i_3}=1,\mathbb X)$ for three arbitrary but different indices $i_1$, $i_2$, and $i_3$. Moreover, equation \eqref{eq:cond_trans} suggests that the conditional probability of node $i_1$ following node $i_3$ given node $i_3$ followed by node $i_2$ and $i_2$ followed by $i_1$ is approximately equal to $\exp(\Delta_{i_3i_2}^\top\beta)/\{2\exp(2\Delta_{i_3i_2}^\top\beta)+1\}^{1/2}$. Let $\pi^{i_1i_2i_3}_{\mathrm{tr}}$ be the approximated conditional probability of $a_{i_1i_3}=1$ given $a_{i_1i_2}a_{i_2i_3}=1$ and $\mathbb X$ as
\begin{align}
\pi^{i_1i_2i_3}_{\mathrm{tr}}=\frac{\gamma_{i_3}}{(\gamma_{i_2}^2+2\gamma_{i_3}^2)^{1/2}}=\frac{\exp(\Delta_{i_3i_2}^\top\beta)}{\{2\exp(2\Delta_{i_3i_2}^\top\beta)+1\}^{1/2}},\label{eq:trans_prob}
\end{align}
which does not depend on $X_{i_1}$. For convenience, we rewrite $\pi^{i_1i_2i_3}_{\mathrm{tr}}$ as $\pi^{i_2i_3}_{\mathrm{tr}}$. Similarly to $\pi_{\mathrm{re}}^{i_1i_2}$, $\pi^{i_2i_3}_{\mathrm{tr}}$ is also free of the nuisance parameter $\alpha_N$ and is strictly larger than 0 as $N\to\infty$. Therefore, we focus on those $(i_1,i_2,i_3)$ triads with $a_{i_1i_2}a_{i_2i_3}=1$ and then treat $a_{i_1i_3}$s as independent binary random variables with response probability $\pi^{i_2i_3}_{\mathrm{tr}}$. This leads to the following loss function
\begin{align}
\mathcal L_{\mathrm{tr}}(\beta) = -\sum_{i_1\neq i_3}\sum_{i_2\neq i_1,i_3}a_{i_1i_2}a_{i_2i_3}\Big\{a_{i_1i_3}\log \pi^{i_2i_3}_{\mathrm{tr}}+\big(1-a_{i_1i_3}\big)\log\big(1-\pi^{i_2i_3}_{\mathrm{tr}}\big)\Big\},\label{eq:liktr}
\end{align}
where the detailed derivation is in Appendix S1.1. Then, a natural and tractable estimator for $\beta$ can be defined as $\widehat \beta_{\mathrm{tr}}=\arg\min_{\beta}\mathcal L_{\mathrm{tr}}(\beta)$ and the asymptotic property of $\widehat\beta_{\mathrm{tr}}$ is given by the following theorem.
\begin{theorem}[Asymptotic Properties of TR]
Assume conditions (C1) and (C2) hold. We then have $N^{(1+\delta)/2}(\widehat\beta_{\mathrm{tr}}-\beta)\xrightarrow[]{d} N(0,C_\alpha^{-1}\mathbf H^{-1}_{\mathrm{tr}}\mathbf M_{\mathrm{tr},1}\mathbf H^{-1}_{\mathrm{tr}})$ if $\delta>0$ and (2) $N^{(1+\delta)/2}(\widehat\beta_{\mathrm{tr}}-\beta)\xrightarrow[]{d} N\{0,C_\alpha^{-1}\mathbf H^{-1}_{\mathrm{tr}}(\mathbf M_{\mathrm{tr},1}+C_\alpha^{-1}\mathbf{M}_{\mathrm{tr},\Delta})\mathbf H^{-1}_{\mathrm{tr}}\}$ if $\delta=0$, where $\mathbf M_{\mathrm{tr},1}$, $\mathbf H_{\mathrm{tr}}$ and $\mathbf{M}_{\mathrm{tr},\Delta}\in\mathbb R^{p\times p}$ are three positive definite matrices defined in Appendix S1.3.
\end{theorem}
\noindent The proof of Theorem 4 is given in Appendix S2.4. By Theorem 4, we find that $\widehat\beta_{\mathrm{tr}}$ is $N^{(1+\delta)/2}$-consistent and asymptotically normal with analytically tractable asymptotic covariance. This is the same convergence rate as that of $\widehat\beta_{\mathrm{re}}$. For valid statistical inference for $\widehat\beta_{\mathrm{tr}}$, $\mathbf M_{\mathrm{tr}}$ and $\mathbf H_{\mathrm{tr}}$ in Theorem 4 need to be consistently estimated. The technical details are given in Appendix S3.1.

{\noindent {\bf Remark 3.} Note that the order of the expected number of terms (i.e., expected computation cost) involved in the corresponding objective function for (1) the PMLE estimator $\widehat\beta_{\rm{pmle}}$ in Section 3.1 is $O(N)$; (2) that for the in-degree estimator $\widehat\beta_{\rm{in}}$ in Section 3.2 is $O(N)$; (3) that for the reciprocity estimator $\widehat\beta_{\rm{re}}$ in Section 3.3 is $O(N^{1+\delta})$; and (4) that for the transitivity estimator $\widehat\beta_{\rm{tr}}$ in Section 3.4 is $O(N^{1+2\delta})$. Our additional simulation results show that the computation costs of $\hat{\beta}_{\text{in}}$ and $\hat{\beta}_{\text{tr}}$ are heavier than those of $\hat{\beta}_{\text{pmle}}$ and $\hat{\beta}_{\text{re}}$, respectively.
Therefore, $\hat{\beta}_{\text{pmle}}$ and $\hat{\beta}_{\text{re}}$ are computationally more preferable than $\hat{\beta}_{\text{in}}$ and $\hat{\beta}_{\text{tr}}$, respectively, if the network size is sufficiently large.}

\csection{SOME EXTENSIONS}
\csubsection{Discussion on Generalized Distribution of Latent Position}

The theories developed in Section~3 rely on an important assumption. That is the normality assumption about the distribution of the latent position $Z_i$. In real practice, however, this assumption could be violated. For example, if different nodes are clustered into different communities, it is then more appropriate to assume for $Z_i$ a distribution with multiple modes.
It is then of great interest to study the sensitivity of those estimators (i.e., $\widehat\beta_{\mathrm{pmle}}$, $\widehat\beta_{\mathrm{in}}$, $\widehat\beta_{\mathrm{re}}$, and $\widehat\beta_{\mathrm{tr}}$) toward the violation of this normality assumption. To this end, we assume in this subsection that  $Z_i\in\mR$ follows a general distribution with a probability density function $f(z)$. Assume $f(z)$ is a sufficiently smooth and bounded function with finite absolute integration for not only itself but also its derivatives of any order. Subsequently, we shall study the asymptotic behaviors of these estimators.

We start with the PMLE estimator $\widehat\beta_{\mathrm{pmle}}$. To this end, we note that the conditional link probability under this general distribution $f(z)$ can be written as
\begin{equation}
E(a_{i_1i_2}\mid \mathbb{X})
= \iint
\exp\!\left\{-\frac{(z_{i_1}-z_{i_2})^2}{2\gamma_{i_2}^2}\right\}
f(z_{i_1})f(z_{i_2})\,dz_{i_1}\,dz_{i_2}
= c_f\,\gamma_{i_2}\{1+o_p(1)\},
\label{eq:netgen}
\end{equation}
where $c_f=\sqrt{2\pi}\|f\|_2^2$ and $\|f\|_s=(\int |f(z)|^sdz)^{1/s}$ for any $s\geq 1$. In case of normally distributed $Z_i$, we have $\|f\|_2=(4\pi)^{-1/4}$ and thus $c_f=1/\sqrt{2}$. Then, \eqref{eq:netgen} becomes the same as \eqref{eq:conprob}. With the help of \eqref{eq:netgen}, the pseudo-log-likelihood can be constructed as 
\begin{equation}
\mathcal L_{\mathrm{pmle}}(\theta)=-\sum_{i=1}^N
\big\{d_i^{\mathrm{in}}(\widetilde{X}_i^\top\theta+\log c_f)
-(N-1-d_i^{\mathrm{in}})\exp(\widetilde{X}_i^\top\theta+\log c_f)\big\}.\label{eq:genlik}
\end{equation}
The verification details of \eqref{eq:netgen} and \eqref{eq:genlik} are given in Appendix S1.5. By~\eqref{eq:genlik}, we can define another intercept parameter as
$\alpha_N^\prime=\alpha_N+\log c_f$.
This implies that the loss function in~\eqref{eq:genlik} becomes basically the same as the original one \eqref{eq:plmelik}. The only difference is that the original intercept $\alpha_N$ in \eqref{eq:plmelik} is now replaced by a new one $\alpha_N^\prime$ in \eqref{eq:genlik}.
This also implies that the loss function ${\mathcal L}_{\mathrm{pmle}}(\theta)$ defined in \eqref{eq:genlik} remains valid in the sense that the resulting estimator~$\widehat\beta_{\mathrm{pmle}}$ is likely to remain consistent for~$\beta$.
However, the resulting estimator~$\widehat\alpha_N$ should be inconsistent, unless the latent position $Z_i$ follows a standard normal distribution.

We next consider the in-degree-based estimator $\widehat\beta_{\mathrm{in}}$. According to \eqref{eq:netgen}, the expected in-degree for node $i$ satisfies
\beqr
E(d_i^{\mathrm{in}}|\mathbb{X})=Nc_f\gamma_i\{1+o_p(1)\}\label{eq:indegreegen}.
\eeqr The verification details in this regard are given in Appendix S1.5. If the latent position $Z_i$ follows a standard normal distribution, we then have $c_f=1/\sqrt{2}$. Then, \eqref{eq:indegreegen} becomes the same as \eqref{eq:in_degree}. With the help of \eqref{eq:indegreegen}, an in-degree-based loss function can then be constructed as
\beqr
\mathcal L_{\mathrm{in}}(\theta)=\sum_{i=1}^N\left\{d_i^{\mathrm{in}}-N\exp(\widetilde X_i^\top\theta+\log c_f)\right\}^2\label{eq:geninlik}
\eeqr
Therefore, the loss function in \eqref{eq:geninlik} becomes the same as that in \eqref{eq:inlik} except that the original intercept $\alpha_N$ is now replaced by a new one $\alpha_N^\prime$ in \eqref{eq:geninlik}. This implies that the original loss function~${\mathcal L}_{\mathrm{in}}(\theta)$ defined in \eqref{eq:inlik} remains valid in the sense that the resulting estimator~$\widehat\beta_{\mathrm{in}}$ is likely to remain consistent for~$\beta$.
However, similar to the previous discussion, the resulting estimator~$\widehat\alpha_N$ is likely to be inconsistent for~$\alpha^\prime_N\neq\alpha_N$, unless $Z_i$ follows a standard normal distribution.

We further consider the reciprocity-based estimator $\widehat\beta_{\mathrm{re}}$. For any two nodes $(i_1,i_2)$, it can be verified that the reciprocity probability $P(a_{i_2i_1}=1| a_{i_1i_2}=1)$  remains the same as \eqref{eq:cond_recip}, even with a general distribution $f(z)$. The verification details are given in Appendix S1.5. This implies that the original loss function defined in \eqref{eq:relik} remains valid in the sense that the resulting estimator~$\widehat\beta_{\mathrm{re}}$ is likely to remain consistent for~$\beta$. Similar story happens for the transitivity-based estimator $\widehat\beta_{\mathrm{tr}}$. Specifically, for any node triplet $(i_1,i_2,i_3)$
and a general $f(z)$, it can be verified that the transitivity probability $P(a_{i_1i_3}=1|a_{i_1i_2}a_{i_2i_3}=1)$ remains to be the same as \eqref{eq:cond_trans}. The verification details are given in Appendix S1.5. As a result, the original loss function defined in \eqref{eq:liktr} remains valid and the resulting estimator~$\widehat\beta_{\mathrm{tr}}$ is likely to remain consistent. To summarize, it seems that all these estimators are likely to remain consistent for $\beta$ but not for $\alpha_N$. Fortunately, in the regression for nodal popularity, the $\beta$-coefficient is the only parameter of primary interest. We then have the following Theorem 5 to confirm this theoretical result formally.

\begin{theorem}[Asymptotics with Non-Normal Distribution]\label{thm:non-normal}
 Assume conditions (C1) and (C2) hold. Further assume that latent position $Z_i$ follows a distribution with general density function $f(z)$. Then, we have, as $N\rightarrow\infty$,
 \begin{itemize}
\item[(1)] $N^{1/2}(\widehat\beta_{\mathrm{pmle}}-\beta)\xrightarrow[]{d} N(0,\widetilde{\mathbb{M}}_{\mathrm{pmle},1})$ if $\delta>0$ and $N^{1/2}(\widehat\beta_{\mathrm{pmle}}-\beta)\xrightarrow[]{d} N(0,\widetilde{\mathbb{M}}_{\mathrm{pmle},0})$ if $\delta=0$.
\item[(2)] $N^{1/2}(\widehat\beta_{\mathrm{in}}-\beta)\xrightarrow[]{d} N(0,\widetilde{\mathbb{M}}_{\mathrm{in},1})$ if $\delta>0$ and $N^{1/2}(\widehat\beta_{\mathrm{in}}-\beta)\xrightarrow[]{d} N(0,\widetilde{\mathbb{M}}_{\mathrm{in},0})$ if $\delta=0$.
\item[(3)] $N^{(1+\delta)/2}(\widehat\beta_{\mathrm{re}}-\beta)\xrightarrow[]{d} N(0,(c_fC_\alpha)^{-1}\widetilde{\mathbf H}^{-1}_{\mathrm{re}}\widetilde{\mathbf M}_{\mathrm{re}}\widetilde{\mathbf H}^{-1}_{\mathrm{re}})$.
\item[(4)] $N^{(1+\delta)/2}(\widehat\beta_{\mathrm{tr}}-\beta)\xrightarrow[]{d} N(0,C_\alpha^{-1}\widetilde{\mathbf H}^{-1}_{\mathrm{tr}}\widetilde{\mathbf M}_{\mathrm{tr},1}\widetilde{\mathbf H}^{-1}_{\mathrm{tr}})$ if $\delta>0$ and $N^{(1+\delta)/2}(\widehat\beta_{\mathrm{tr}}-\beta)\xrightarrow[]{d} N\{0,C_\alpha^{-1}\widetilde{\mathbf H}^{-1}_{\mathrm{tr}}(\widetilde{\mathbf M}_{\mathrm{tr},1}+C_\alpha^{-1}\widetilde{\mathbf M}_{\mathrm{tr},\Delta})\widetilde{\mathbf H}^{-1}_{\mathrm{tr}}\}$ if $\delta=0$.
 \end{itemize}

\end{theorem}

\noindent The analytical details for the quantities (i.e., $\widetilde{\mathbb{M}}_{\mathrm{pmle},0}$, $\widetilde{\mathbb{M}}_{\mathrm{pmle},1}$, $\widetilde{\mathbb{M}}_{\mathrm{in},0}$, $\widetilde{\mathbb{M}}_{\mathrm{in},1}$, $\widetilde{\mathbf H}_{\mathrm{re}}$, $\widetilde{\mathbf M}_{\mathrm{re}}$, $\widetilde{\mathbf H}_{\mathrm{tr}}$, $\widetilde{\mathbf M}_{\mathrm{tr},1}$, and $\widetilde{\mathbf M}_{\mathrm{tr},\Delta}$) are given in Appendix S1.5 and the rigorous technical proofs for Theorem 5 are given in Appendix S2.5. By Theorem 5, we know that the four estimators (i.e., $\widehat\beta_{\mathrm{pmle}}$, $\widehat\beta_{\mathrm{in}}$, $\widehat\beta_{\mathrm{re}}$, and $\widehat\beta_{\mathrm{tr}}$) remain to be consistent with the same rates as given in Theorems 1--4, respectively. Meanwhile, they remain to be asymptotically normal but with asymptotic covariance matrices somewhat different from those under the normality assumption. If $f(z)$ happens to be a standard normal distribution, we then have $\|f\|_1=1$, $\|f\|_2=(4\pi)^{-1/4}$, and $\|f\|_3=(2\pi\sqrt{3})^{-1/3}$. Then the results given in Theorem 5 reduce to their corresponding counterparts given in Theorems 1--4.

{\csubsection{Estimation based on Higher-Order Statistics}

Our previous investigation reveals that the estimators based on transitivity $\widehat\beta_{\mathrm{tr}}$  and reciprocity $\widehat\beta_{\mathrm{re}}$ exhibit faster convergence rates than those based on the pseudo likelihood  $\widehat\beta_{\mathrm{pmle}}$ and in-degree $\widehat\beta_{\mathrm{in}}$. This naturally raises the question of whether even higher-order network statistics can lead to even faster convergence rates. To explore this possibility, we proceed to study one particular type of higher-order statistics here. Those are the common neighbor statistics.  Empirically, it has been well documented that the presence of multiple common neighbors between a pair of nodes leads to a higher chance of a direct tie.

Consider two arbitrary but distinct nodes $i_1$ and $i_2$.
Let $\{k_1,\ldots,k_m\}$ be a set of $m$ common neighbors of $(i_1,i_2)$ with
$a_{i_1 k_\ell}=a_{k_\ell i_2}=1$, for $\ell=1,\ldots,m.$
It is then of great interest to evaluate $\mathcal{P}_{\mathrm{cn}}=P(a_{i_1i_2}=1|\prod_{\ell=1}^m a_{i_1 k_\ell} a_{k_\ell i_2}=1,\mX)$; see \cite{chang2019popularity}.  This leads to an analytical solution as,
\begin{equation}
\label{eq:cn-limit-sigma}
\mathcal{P}_{\mathrm{cn}}
=
\left\{
\sum_{\ell=1}^m(\gamma_{k_\ell}^2+\gamma_{i_2}^2)^{-1}\right\}
\left\{\gamma_{i_2}^{-2}+\sum_{\ell=1}^m(\gamma_{k_\ell}^2+\gamma_{i_2}^2)^{-1}\right\}^{-1/2}
+ \exp(2\alpha_N)O_p\big(1\big).
\end{equation}
In the case of $m=1$, this probability reduces to \eqref{eq:trans_prob} for transitivity. For illustration purposes, we consider a slightly more complicated case with $m=2$.
Then by \eqref{eq:cn-limit-sigma}, we have $P(a_{i_1i_2}=1|\prod_{\ell=1}^2 a_{i_1 k_\ell} a_{k_\ell i_2}=1,\mX)=\pi_{\mathrm{cn}}^{i_1i_2 k_1 k_2}\{1+o_p(1)\}$, with
\begin{equation}
\label{eq:cn-pi}
\pi_{\mathrm{cn}}^{i_1i_2 k_1 k_2}
=\left\{
s(\Delta_{k_1 i_2}^\top\beta)+s(\Delta_{k_2 i_2}^\top\beta)\right\}\left\{1+s(\Delta_{k_1 i_2}^\top\beta)+s(\Delta_{k_2 i_2}^\top\beta)
\right\}^{-1/2},
\end{equation}
where $s(x)=(1+e^{2x})^{-1}$, and $s(\Delta_{k_\ell i_2}^\top\beta)=
\big\{1+\exp(2\Delta_{k_\ell i_2}^\top\beta)\big\}^{-1}=\gamma_{i_2}^2(\gamma_{k_\ell}^2+\gamma_{i_2}^2)^{-1}$.
For convenience, we rewrite $\pi_{\mathrm{cn}}^{i_1i_2 k_1 k_2}$ as $\pi^{i_2k_1 k_2}_{\mathrm{cn}}$. Similar to the likelihood based on transitivity, a pseudo-log-likelihood type loss function can be constructed as
\begin{align*}
\mathcal L_{\mathrm{cn}}(\beta)
=-\sum_{i_1\neq i_2}
\sum_{\substack{k_1\neq k_2\\ k_1,k_2\notin\{i_1,i_2\}}}
a_{i_1i_2k_1k_2}\n\Big\{a_{i_1i_2}\log \pi_{\mathrm{cn}}^{i_2 k_1 k_2}+\big(1-a_{i_1 i_2}\big)\log\big(1-\pi_{\mathrm{cn}}^{i_2 k_1 k_2}\big)
\Big\},
\end{align*}
where $a_{i_1i_2k_1k_2}=a_{i_1 k_1}a_{k_1 i_2}\,a_{i_1 k_2}a_{k_2 i_2}$. This leads to a common-neighbor based (CN) estimator as $\widehat\beta_{\mathrm{cn}}=\arg\min_\beta\mathcal L_{\mathrm{cn}}(\beta)$, whose asymptotic properties are given by Theorem 6 presented as follows.
\begin{theorem}[Asymptotic Properties of CN]
 Assume conditions (C1) and (C2) hold. We have $N^{(1+\delta)/2}(\widehat\beta_{\mathrm{cn}}-\beta)\xrightarrow[]{d} N(0,C_\alpha^{-1}\mathbf H^{-1}_{\mathrm{cn}}\mathbf M_{\mathrm{cn},1}\mathbf H^{-1}_{\mathrm{cn}})$ if $\delta>0$ and (2) $N^{(1+\delta)/2}(\widehat\beta_{\mathrm{cn}}-\beta)\xrightarrow[]{d} N\{0,C_\alpha^{-1}\mathbf H^{-1}_{\mathrm{cn}}(\mathbf M_{\mathrm{cn},1}+C_\alpha^{-1}\mathbf{M}_{\mathrm{cn},\Delta})\mathbf H^{-1}_{\mathrm{cn}}\}$ if $\delta=0$, where $\mathbf M_{\mathrm{cn},1}\in\mathbb R^{p\times p}$, $\mathbf{M}_{\mathrm{cn},\Delta}\in\mathbb R^{p\times p}$, and $\mathbf H_{\mathrm{cn}}\in\mathbb R^{p\times p}$ are three positive definite matrices defined in Appendix S1.4.
 \end{theorem}
\noindent Similar to Theorem 4, we find that $\widehat\beta_{\mathrm{cn}}$ is $N^{(1+\delta)/2}$-consistent and asymptotically normal with different asymptotic covariances
according to the value of density level $\delta$. This is the convergence rate remains to be the same as that of $\widehat\beta_{\mathrm{re}}$ and $\widehat\beta_{\mathrm{tr}}$. Therefore, it seems that exploring higher-order statistics does not necessarily lead to a faster convergence rate. This interesting finding is not totally surprising. If we treat each edge as an ``effective sample'', then the total ``effective sample size'' of the whole network is of the order $O(N^{1+\delta})$. Accordingly, no estimator can have a convergence rate faster than $O_p(N^{(1+\delta)/2})$, which has already been achieved by $\hat\beta_{\rm{re}}$ and $\hat\beta_{\rm{tr}}$.

\csection{NUMERICAL STUDY}
\csubsection{Simulation Studies}

To demonstrate the finite sample performance of the proposed estimators for PoRe-LSM, we conduct a number of simulation studies in this subsection. Specifically, we have $Z_i$s independently generated from a standard normal distribution. Next, we generate $X_i=(X_{ij})\in\mathbb R^5$ from a multivariate normal distribution with mean $0$ and covariance matrix $\Sigma_X=(\sigma_{X,j_1j_2})$, where $\sigma_{X,j_1j_2}=0.5^{|j_1-j_2|}$. The true regression coefficient is set to $\beta = (-0.2, 0.2, -0.1, 0.1, 0)^\top$ and the intercept parameter is set to $\alpha_N = \log C_\alpha -(1-\delta)\log N$ with $C_\alpha = 15$. Different network sizes $N\in\{5000,10000,20000,30000\}$ and density levels $\delta\in\{0,0.25\}$ are considered. Since $\theta = (\theta_j)=(\alpha_N,\beta^\top)^\top\in\mathbb R^6$ and $\widetilde X_i=(1,X_i^\top)^\top\in\mathbb R^6$, we next compute $\gamma_{i}$ as $\exp(\widetilde X_i^\top\theta)$ according to \eqref{eq:pop_reg}. Therefore, $a_{i_1i_2}$s are independently generated from the Bernoulli distribution according to the conditional probability defined in \eqref{eq:latent_model}. Thereafter, various estimators (i.e., $\widehat\beta_{\mathrm{pmle}}$, $\widehat\beta_{\mathrm{in}}$, $\widehat\beta_{\mathrm{re}}$, and $\widehat\beta_{\mathrm{tr}}$) can be calculated, and their finite sample performances are then evaluated as follows.

Specifically, with a given parameter setting (i.e., $N$, $p$, $\Sigma_X$, $\beta$, $C_\alpha$, and $\delta$), we randomly replicate the experiment $B=1000$ times. Denote the estimator obtained in the $b$th replication $\widehat\beta^{(b)}=(\widehat\beta^{(b)}_j)\in\mathbb R^5$ with $1\leq b\leq B$. We then compute the root mean square error (RMSE) for each coefficient estimator as $\text{RMSE}_j=\{B^{-1}\sum_{b=1}^B(\widehat\beta^{(b)}_j-\beta_j)^2\}^{1/2}$. Moreover, we compute the corresponding Monte Carlo standard error as ${\text{SE}}_j=\sqrt{B^{-1}\sum_{b=1}^B(\widehat \beta_{j}^{(b)}-\bar\beta_{j})^2}$, where $\bar\beta_{j}=B^{-1}\sum_{b=1}^B\widehat \beta_{j}^{(b)}$. On the other hand, the covariance matrix can be estimated according to the formula derived for each replication, e.g., $\widehat{\text{cov}}(\widehat\beta_{\mathrm{re}})$ as $\widehat{\mathbf{H}}_{\mathrm{re}}^{-1}\widehat{\mathbf{M}}_{\mathrm{re}}\widehat{\mathbf{H}}_{\mathrm{re}}^{-1}/\{N^2\exp(\widehat\alpha_\mathrm{re})\}$. Then, denote the estimated standard error of $\widehat\beta^{(b)}$ as $\widehat{\text{SE}}^{(b)}_j$. To evaluate the performance of the estimated standard error, we define the absolute relative error (ARE) for $\widehat{\text{SE}}^{(b)}_j$ as $\text{ARE}_j=B^{-1}\sum_{b=1}^B |\widehat{\text{SE}}^{(b)}_j/{\text{SE}}_j-1|\times 100\%$. Based on $\widehat\beta^{(b)}_j$ and $\widehat{\text{SE}}^{(b)}_j$, a $95\%$ confidence interval can be constructed as $\text{CI}^{(b)}_j=(\widehat \beta_{j}^{(b)}-z_{0.975}\widehat{\text{SE}}^{(b)}_j,\widehat \beta_{j}^{(b)}+z_{0.975}\widehat{\text{SE}}^{(b)}_j)$, where $z_\alpha$ is the $\alpha$-th lower quantile of the standard normal distribution. Finally, the empirical coverage probability (ECP) can be calculated as $\text{ECP}_{j}=B^{-1}\sum_{b=1}^B I\{\beta_j\in\text{CI}^{(b)}_j\}$, where $I(\cdot)$ is the indicator function.

For each estimator, define $\overline{\text{RMSE}}=\sum_{j=1}^p\text{RMSE}_j$, $\overline{\text{ARE}}=\sum_{j=1}^p\text{ARE}_j$, and $\overline{\text{ECP}}=\sum_{j=1}^p\text{ECP}_j$ as the average value of measures for all dimensions of $\hat{\beta}$. The simulation results of $\overline{\text{RMSE}}$, $\overline{\text{ARE}}$, and $\overline{\text{ECP}}$ are summarized in Table \ref{tb_simu}. In case of $\delta=0$, we find that the RMSE values steadily decrease toward 0 as $N$ increases for all estimators (i.e., $\widehat\beta_{\mathrm{pmle}}$, $\widehat\beta_{\mathrm{in}}$, $\widehat\beta_{\mathrm{re}}$, and $\widehat\beta_{\mathrm{tr}}$), confirming the consistency results given in Theorems 1--4. Most ARE values of $\widehat{\text{SE}}$ are close to 0, indicating that the estimated standard errors closely match their Monte Carlo counterparts. The ECP values are all around the nominal level 95\%, supporting the asymptotic normality of the estimators. The findings for $\delta=0.25$ are largely similar but with one key difference. That is the performance difference in RMSE between the first-order estimators ($\widehat\beta_{\mathrm{pmle}}$, $\widehat\beta_{\mathrm{in}}$) and the higher-order estimators ($\widehat\beta_{\mathrm{re}}$, $\widehat\beta_{\mathrm{tr}}$) is more apparent. This is mainly because the convergence rates of $\widehat\beta_{\mathrm{pmle}}$ and $\widehat\beta_{\mathrm{in}}$ remain $N^{1/2}$, whereas those of $\widehat\beta_{\mathrm{re}}$ and $\widehat\beta_{\mathrm{tr}}$ are $N^{(1+\delta)/2}$ by our Theorems 3 and 4. More detailed simulation results about each dimension $\widehat \beta_j$ are given in Tables 1 and 2 in Appendix S4.1.

\begin{table*}[h!]\footnotesize
\renewcommand{\arraystretch}{1}
\setlength{\tabcolsep}{2.5pt}
\centering
\centering
\begin{tabular}{c|c|c|c|c|c|c|c}
\hline
\multicolumn{1}{c|}{\multirow{2}{*}{$N$}}&\multicolumn{1}{c|}{\multirow{2}{*}{Estimators}}& \multicolumn{1}{c}{} & \multicolumn{1}{c}{$\delta=0$} & \multicolumn{1}{c|}{}&\multicolumn{1}{c}{} & \multicolumn{1}{c}{$\delta=0.25$} & \multicolumn{1}{c}{}\\
& &\multicolumn{1}{c}{$\overline{\text{RMSE}}\text{(\textperthousand)}$} & \multicolumn{1}{c}{$\overline{\text{ARE}}\text{(\%)}$} & \multicolumn{1}{c|}{$\overline{\text{ECP}}\text{(\%)}$}& \multicolumn{1}{c}{$\overline{\text{RMSE}}\text{(\textperthousand)}$} & \multicolumn{1}{c}{$\overline{\text{ARE}}\text{(\%)}$} & \multicolumn{1}{c}{$\overline{\text{ECP}}\text{(\%)}$}\\
\hline
\multirow{4}{*}{5000}& PMLE &
\multicolumn{1}{c}{8.93} & \multicolumn{1}{c}{2.41} & \multicolumn{1}{c|}{94.7}&\multicolumn{1}{c}{7.80} & \multicolumn{1}{c}{3.37} & \multicolumn{1}{c}{92.5}\\
& IN &
\multicolumn{1}{c}{9.50} & \multicolumn{1}{c}{2.73} & \multicolumn{1}{c|}{94.8}&\multicolumn{1}{c}{7.83} & \multicolumn{1}{c}{1.91} & \multicolumn{1}{c}{94.9}\\
& RE &
\multicolumn{1}{c}{4.27} & \multicolumn{1}{c}{1.71} & \multicolumn{1}{c|}{95.3}&\multicolumn{1}{c}{1.47} & \multicolumn{1}{c}{1.79} & \multicolumn{1}{c}{95.3}\\
& TR &\multicolumn{1}{c}{5.27} & \multicolumn{1}{c}{3.08} & \multicolumn{1}{c|}{94.2}&\multicolumn{1}{c}{1.61} & \multicolumn{1}{c}{3.07} & \multicolumn{1}{c}{94.2}\\
\hline
\multirow{4}{*}{10000}& PMLE &
\multicolumn{1}{c}{6.41} & \multicolumn{1}{c}{2.20} & \multicolumn{1}{c|}{94.6}&\multicolumn{1}{c}{5.46} & \multicolumn{1}{c}{4.51} & \multicolumn{1}{c}{92.9} \\
&IN &\multicolumn{1}{c}{6.82} & \multicolumn{1}{c}{1.96} & \multicolumn{1}{c|}{94.8}&\multicolumn{1}{c}{5.64} & \multicolumn{1}{c}{3.27} & \multicolumn{1}{c}{94.1}\\
& RE &
\multicolumn{1}{c}{3.04} & \multicolumn{1}{c}{1.52} & \multicolumn{1}{c|}{95.1}&\multicolumn{1}{c}{0.95} & \multicolumn{1}{c}{1.87} & \multicolumn{1}{c}{95.2}\\
& TR &\multicolumn{1}{c}{3.59} & \multicolumn{1}{c}{2.24} & \multicolumn{1}{c|}{95.6}&\multicolumn{1}{c}{1.01} & \multicolumn{1}{c}{1.32} & \multicolumn{1}{c}{95.4}\\
\hline
\multirow{4}{*}{20000}& PMLE &
\multicolumn{1}{c}{4.41} & \multicolumn{1}{c}{0.85} & \multicolumn{1}{c|}{95.2}&\multicolumn{1}{c}{3.76} & \multicolumn{1}{c}{2.32} & \multicolumn{1}{c}{93.5} \\
&IN &\multicolumn{1}{c}{4.72} & \multicolumn{1}{c}{1.00} & \multicolumn{1}{c|}{95.2}&\multicolumn{1}{c}{3.90} & \multicolumn{1}{c}{1.42} & \multicolumn{1}{c}{95.0}\\
& RE &
\multicolumn{1}{c}{2.15} & \multicolumn{1}{c}{1.35} & \multicolumn{1}{c|}{95.0}&\multicolumn{1}{c}{0.63} & \multicolumn{1}{c}{2.47} & \multicolumn{1}{c}{95.1}\\
& TR &\multicolumn{1}{c}{2.62} & \multicolumn{1}{c}{2.56} & \multicolumn{1}{c|}{94.4}&\multicolumn{1}{c}{0.67} & \multicolumn{1}{c}{2.14} & \multicolumn{1}{c}{94.8}\\
\hline
\multirow{4}{*}{30000}& PMLE &
\multicolumn{1}{c}{3.69} & \multicolumn{1}{c}{1.91} & \multicolumn{1}{c|}{94.3}&\multicolumn{1}{c}{3.01} & \multicolumn{1}{c}{2.27} & \multicolumn{1}{c}{94.1} \\
&IN &\multicolumn{1}{c}{3.93} & \multicolumn{1}{c}{1.65} & \multicolumn{1}{c|}{94.3}&\multicolumn{1}{c}{3.18} & \multicolumn{1}{c}{1.59} & \multicolumn{1}{c}{94.7}\\
& RE &
\multicolumn{1}{c}{1.73} & \multicolumn{1}{c}{1.61} & \multicolumn{1}{c|}{95.4}&\multicolumn{1}{c}{0.48} & \multicolumn{1}{c}{1.95} & \multicolumn{1}{c}{95.4}\\
& TR &\multicolumn{1}{c}{2.13} & \multicolumn{1}{c}{1.66} & \multicolumn{1}{c|}{94.8}&\multicolumn{1}{c}{0.50} & \multicolumn{1}{c}{2.16} & \multicolumn{1}{c}{95.6}\\
\hline
\end{tabular}
\caption{The simulation results for all estimators in PoRe-LSM. The number of replications is $B=1000$. The average root mean square error $\overline{\text{RMSE(\textperthousand)}}$, absolute relative error $\overline{\text{ARE(\%)}}$, and empirical coverage probability $\overline{\text{ECP(\%)}}$ are reported with different network sizes $N$ and density levels $\delta$.}
\label{tb_simu}
\end{table*}

\csubsection{Real Data Analysis}

In this subsection, we present a real data example. Specifically, we use a large-scale statistician author citation network collected by \cite{GAO2024121634}. Each node in this data represents an author. Those isolated nodes and nodes with very large degrees are excluded. This leaves the final dataset with 36,210 nodes. A citation network can be constructed accordingly. Define $a_{i_1i_2}=1$ if the $i_1$th author has cited at least one paper of the $i_2$th author and $a_{i_1i_2}=0$ otherwise, and set $a_{ii}=0$ for $1\leq i \leq N$.

Next, we collect for each node a total of three features. The first feature is the $h$-index (HID) $X_{i1}$, which is defined to be the 3-year lagged $h$-index of author $i$. The second feature is the average topic (ATD) $X_{i2}$, which is defined to be the average topic diversity of all papers by author $i$ (ATD) $X_{i2}$. Here, the topic diversity of a paper is the entropy of the paper's topic distribution. The third feature is the in-degree of the node (IND) $X_{i3}$.  To avoid endogeneity introduced by IND, we randomly split the network into two parts by node with equal size: a feature network with node index set $\mathcal{D}_{\rm{feature}}$ and a main network with node index set $\mathcal{D}_{\rm{main}}$. Therefore, for $i\in\mathcal{D}_{\rm{main}}$, $X_{i3}$ is computed as the number of followers of node $i$ in $\mathcal{D}_{\rm{feature}}$. All features are $\log(x+1)$-transformed before the analysis. We next compute the four estimators ($\widehat\beta_{\mathrm{pmle}}$, $\widehat\beta_{\mathrm{in}}$, $\widehat\beta_{\mathrm{re}}$ and $\widehat\beta_{\mathrm{tr}}$) for the main network, together with their $\widehat{\text{SE}}$, $p$-values and the confidence intervals (CI). The detailed results are reported in Table 8 in Appendix S4.7. We find that HID and IND are significantly positive at the $5\%$ level of significance for all methods, suggesting an author’s popularity is closely related to the influence of the author’s work and the number of authors who have cited the author’s work.

We next demonstrate the performance of our model for link prediction, an important task in network analysis \citep{link2007,lu2011link}. We follow the edge cross-validation (ECV) strategy proposed by \cite{li2020network}.
Specifically, a subset of edges is randomly selected as the ``held-out set'' for testing. The fraction of subsampled edges from $\mathcal{D}_{\rm{main}}$ for the held-out set is set to be only 10\%. Therefore, we can treat the network constructed by the remaining edges as if the network were complete. This constitutes our training network. We then compute the four estimators ($\widehat\beta_{\mathrm{pmle}}$, $\widehat\beta_{\mathrm{in}}$, $\widehat\beta_{\mathrm{re}}$ and $\widehat\beta_{\mathrm{tr}}$) on this training network and estimate the popularity of each node in the training network according to \eqref{eq:pop_reg}. Next, by \eqref{eq:cn-limit-sigma}, we estimate the following conditional probability for all entities in the held-out set as,
\begin{align*}                                       \widehat P\left(a_{i_1i_2}=1\Bigg|\prod_{i_3=1}^{m}a_{i_1i_3}a_{i_3i_2}=1,\mathbb X\right)=\left\{\frac{\sum_{i_3=1}^{m}(\widehat\gamma^2_{i_3}+\widehat\gamma^2_{i_2})^{-1}}{\widehat\gamma^{-2}_{i_2}+\sum_{i_3=1}^{m}(\widehat\gamma^2_{i_3}+\widehat\gamma^2_{i_2})^{-1}}\right\}^{1/2}
\end{align*}
where $m$ is the number of common neighbors of $i_1$ and $i_2$ in training network and $i_3$s are common neighbors such that $a_{i_1i_3}a_{i_3i_2}=1$.
In case no common neighbor in training network is shared, we directly estimate the conditional probability as $\widehat P\left(a_{i_1i_2}=1|\mathbb X\right)=\widehat{\gamma}_{i_2}/\sqrt{2}$. Then, the receiver operating characteristic (ROC) curve can be obtained and the area under ROC curve (AUC) can be computed. Each experiment is repeated 100 times and this leads to a total of 100 ROC curves and AUC values.

For comparison purposes, a number of classical methods for link prediction are also evaluated \citep{link2007,lu2011link}. These methods are Common Neighbors Index (CNI), Salton Index (SI), Sørensen Index (SOI), Hub Promoted Index (HPI), Hub Depressed Index (HDI), Leicht-Holme-Newman Index (LHNI), Adamic-Adar Index (AAI), and Resource Allocation Index (RAI). In addition, we include the probability index (PI) of \cite{chang2019popularity} and ECV of \cite{li2020network}. Here, the PI estimator is a simple moment-based estimator $\widehat\gamma_i = \sqrt{2} d_i^{\mathrm{in}}/N$ without taking any covariate information into consideration. See Appendix S4.7 for more details. We find that all the methods proposed in this work consistently achieve higher AUC values than the other competing benchmarks. The detailed results are given in Table \ref{tb_AUC} and the averaged ROC curves are presented in Figure 3 in Appendix S4.7. Specifically, averaged AUC values of PMLE, IN, RE, and TR are all larger than 82.0\%. The PI method ranks as the fifth-best method with AUC 79.9\% and ECV is the sixth-best one with AUC 79.2\%.
\begin{table*}[h!]\footnotesize
\renewcommand{\arraystretch}{1.5}
\setlength{\tabcolsep}{3.75pt}
\centering
\begin{tabular}{c|ccccccc}
\hline
Methods & PMLE & IN & RE & TR & PI & ECV & CNI \\
\hline
AUC values & 0.830 & 0.829 & 0.829 & 0.827 & 0.799 & 0.792 & 0.619 \\
\hline
Methods & SI & SOI & HPI & HDI & LHNI & AAI & RAI \\
\hline
AUC values & 0.619 & 0.619 & 0.619 & 0.619 & 0.619 & 0.624 & 0.624 \\
\hline
\end{tabular}
\caption{The averaged AUC results of the PoRe-LSM and other methods with number of replications $B = 100$.}
\label{tb_AUC}
\end{table*}

\csection{CONCLUDING REMARKS}

To conclude the article, we discuss here a number of interesting topics for future research. First, this study focuses on a single-layer and single-mode network. This is not suitable for networks with multiple types of relationships (e.g., multiple social platforms) and multiple types of nodes (e.g., restaurant-customer network). Therefore, the PoRe-LSM for more complex networks (e.g., multilayer network, bipartite network) is an interesting topic to study \citep{friel2016interlocking,zhang2020flexible}. Second, this study considers a fixed-dimensional nodal feature. In real applications, we might encounter a network with high-dimensional nodal features. Then, how to conduct the popularity regression with high-dimensional features is another interesting problem for future study \citep{zhang2022joint}. Lastly, the log-linear regression relationship assumed in the PoRe-LSM could be too restrictive. Therefore, a more flexible semiparametric model (e.g., a partially log-linear model) might also be a worthwhile in-depth study.

\noindent
\textbf{Acknowledgements.} Danyang Huang's research is supported by the National Natural Science Foundation of China (72471230), the MOE Project of Key Research Institute of Humanities and Social Sciences (22JJD110001), Chinese institute for public governance research, Big Data and Responsible Artificial Intelligence for National Governance, Renmin University of China. Rui Pan's research is supported by the National Natural Science Foundation of China (No. 72471254). Hansheng Wang's research is partially supported by the National Natural Science Foundation of China (72621002, 72495123, 12271012).

\noindent
\textbf{Conflict of Interest Statement.} The authors report there are no competing interests to declare.
\begin{center}
\bibliographystyle{asa}
\bibliography{ref}

@article{hoff2002latent,
  title={Latent Space Approaches to Social Network Analysis},
  author={Hoff, Peter D and Raftery, Adrian E and Handcock, Mark S},
  journal={Journal of the American Statistical Association},
  volume={97},
  number={460},
  pages={1090--1098},
  year={2002},
  publisher={Taylor \& Francis}
}

@article{chang2019popularity,
  title={A Popularity-Scaled Latent Space Model for Large-Scale Directed Social Network},
  author={Chang, Xiangyu and Huang, Danyang and Wang, Hansheng},
  journal={Statistica Sinica},
  volume={29},
  number={3},
  pages={1277--1299},
  year={2019},
  publisher={JSTOR}
}

@article{watts1998collective,
  title={Collective Dynamics of Small-World Networks},
  author={Watts, Duncan J and Strogatz, Steven H},
  journal={Nature},
  volume={393},
  number={6684},
  pages={440--442},
  year={1998},
  publisher={Nature Publishing Group}
}

@article{schweinberger2015local,
  title={Local Dependence in Random Graph Models: Characterization, Properties and Statistical Inference},
  author={Schweinberger, Michael and Handcock, Mark S},
  journal={Journal of the Royal Statistical Society Series B: Statistical Methodology},
  volume={77},
  number={3},
  pages={647--676},
  year={2015},
  publisher={Oxford University Press}
}

@inproceedings{wang2020logistic,
  title={Logistic Regression for Massive Data With Rare Events},
  author={Wang, HaiYing},
  booktitle={International Conference on Machine Learning},
  pages={9829--9836},
  year={2020},
  organization={PMLR}
}

@article{sewell2015latent,
  title={Latent Space Models for Dynamic Networks},
  author={Sewell, Daniel K and Chen, Yuguo},
  journal={Journal of the American Statistical Association},
  volume={110},
  number={512},
  pages={1646--1657},
  year={2015},
  publisher={Taylor \& Francis}
}

@article{holland1981exponential,
  title={An Exponential Family of Probability Distributions for Directed Graphs},
  author={Holland, Paul W and Leinhardt, Samuel},
  journal={Journal of the American Statistical Association},
  volume={76},
  number={373},
  pages={33--50},
  year={1981},
  publisher={Taylor \& Francis}
}

@article{strauss1990pseudolikelihood,
  title={Pseudolikelihood Estimation for Social Networks},
  author={Strauss, David and Ikeda, Michael},
  journal={Journal of the American Statistical Association},
  volume={85},
  number={409},
  pages={204--212},
  year={1990},
  publisher={Taylor \& Francis}
}

@article{frank1986markov,
  title={Markov Graphs},
  author={Frank, Ove and Strauss, David},
  journal={Journal of the American Statistical Association},
  volume={81},
  number={395},
  pages={832--842},
  year={1986},
  publisher={Taylor \& Francis}
}

@article{wang1987stochastic,
  title={Stochastic Blockmodels for Directed Graphs},
  author={Wang, Yuchung J and Wong, George Y},
  journal={Journal of the American Statistical Association},
  volume={82},
  number={397},
  pages={8--19},
  year={1987},
  publisher={Taylor \& Francis}
}

@article{holland1983stochastic,
  title={Stochastic Blockmodels: First Steps},
  author={Holland, Paul W and Laskey, Kathryn Blackmond and Leinhardt, Samuel},
  journal={Social Networks},
  volume={5},
  number={2},
  pages={109--137},
  year={1983},
  publisher={Elsevier}
}

@article{macdonald2022latent,
  title={Latent Space Models for Multiplex Networks With Shared Structure},
  author={MacDonald, Peter W and Levina, Elizaveta and Zhu, Ji},
  journal={Biometrika},
  volume={109},
  number={3},
  pages={683--706},
  year={2022},
  publisher={Oxford University Press}
}

@article{zhang2022joint,
  title={Joint Latent Space Models for Network Data With High-Dimensional Node Variables},
  author={Zhang, Xuefei and Xu, Gongjun and Zhu, Ji},
  journal={Biometrika},
  volume={109},
  number={3},
  pages={707--720},
  year={2022},
  publisher={Oxford University Press}
}

@article{aoas2016Ji,
author = {Pengsheng Ji and Jiashun Jin},
title = {{Coauthorship and Citation Networks for Statisticians}},
volume = {10},
journal = {The Annals of Applied Statistics},
number = {4},
publisher = {Institute of Mathematical Statistics},
pages = {1779 -- 1812},
year = {2016},
}

@article{friel2016interlocking,
  title={Interlocking Directorates in Irish Companies Using a Latent Space Model for Bipartite Networks},
  author={Friel, Nial and Rastelli, Riccardo and Wyse, Jason and Raftery, Adrian E},
  journal={Proceedings of the National Academy of Sciences},
  volume={113},
  number={24},
  pages={6629--6634},
  year={2016},
  publisher={National Acad Sciences}
}

@article{chun2015gene,
  title={Gene Regulation Network Inference With Joint Sparse Gaussian Graphical Models},
  author={Chun, Hyonho and Zhang, Xianghua and Zhao, Hongyu},
  journal={Journal of Computational and Graphical Statistics},
  volume={24},
  number={4},
  pages={954--974},
  year={2015},
  publisher={Taylor \& Francis}
}

@inproceedings{zhang2020flexible,
  title={A Flexible Latent Space Model for Multilayer Networks},
  author={Zhang, Xuefei and Xue, Songkai and Zhu, Ji},
  booktitle={International Conference on Machine Learning},
  pages={11288--11297},
  year={2020},
  organization={PMLR}
}

@article{liu2022variational,
  title={Variational Inference for Latent Space Models for Dynamic Networks},
  author={Liu, Yan and Chen, Yuguo},
  journal={Statistica Sinica},
  volume={32},
  number={4},
  pages={2147--2170},
  year={2022},
  publisher={JSTOR}
}

@article{zhao2024structured,
  title={Structured Optimal Variational Inference for Dynamic Latent Space Models},
  author={Zhao, Peng and Bhattacharya, Anirban and Pati, Debdeep and Mallick, Bani K},
  journal={Journal of Machine Learning Research},
  volume={25},
  number={259},
  pages={1--55},
  year={2024}
}

@article{hunter2008goodness,
  title={Goodness of Fit of Social Network Models},
  author={Hunter, David R and Goodreau, Steven M and Handcock, Mark S},
  journal={Journal of the American Statistical Association},
  volume={103},
  number={481},
  pages={248--258},
  year={2008},
  publisher={Taylor \& Francis}
}

@article{ji2022co,
  title={Co-citation and Co-authorship Networks of Statisticians},
  author={Ji, Pengsheng and Jin, Jiashun and Ke, Zheng Tracy and Li, Wanshan},
  journal={Journal of Business \& Economic Statistics},
  volume={40},
  number={2},
  pages={469--485},
  year={2022},
  publisher={Taylor \& Francis}
}

@article{handcock2007model,
  title={Model-Based Clustering for Social Networks},
  author={Handcock, Mark S and Raftery, Adrian E and Tantrum, Jeremy M},
  journal={Journal of the Royal Statistical Society Series A: Statistics in Society},
  volume={170},
  number={2},
  pages={301--354},
  year={2007},
  publisher={Oxford University Press}
}

@article{wu2020comprehensive,
  title={A Comprehensive Survey on Graph Neural Networks},
  author={Wu, Zonghan and Pan, Shirui and Chen, Fengwen and Long, Guodong and Zhang, Chengqi and Philip, S Yu},
  journal={IEEE Transactions on Neural Networks and Learning Systems},
  volume={32},
  number={1},
  pages={4--24},
  year={2020},
  publisher={IEEE}
}

@inproceedings{young2007random,
  title={Random Dot Product Graph Models for Social Networks},
  author={Young, Stephen J and Scheinerman, Edward R},
  booktitle={International Workshop on Algorithms and Models for the Web-Graph},
  pages={138--149},
  year={2007},
  organization={Springer}
}

@article{barabasi1999emergence,
  title={Emergence of Scaling in Random Networks},
  author={Barab{\'a}si, Albert-L{\'a}szl{\'o} and Albert, R{\'e}ka},
  journal={Science},
  volume={286},
  number={5439},
  pages={509--512},
  year={1999},
  publisher={American Association for the Advancement of Science}
}

@article{karrer2011stochastic,
  title={Stochastic Blockmodels and Community Structure in Networks},
  author={Karrer, Brian and Newman, Mark EJ},
  journal={Physical Review E—Statistical, Nonlinear, and Soft Matter Physics},
  volume={83},
  number={1},
  pages={016107},
  year={2011},
  publisher={APS}
}

@article{Weilan2021influence,
 ISSN = {10170405, 19968507},
 URL = {https://www.jstor.org/stable/27089314},
 author = {Tao Zou and Ronghua Luo and Wei Lan and Chih-Ling Tsai},
 journal = {Statistica Sinica},
 number = {4},
 pages = {pp. 1727--1748},
 publisher = {Institute of Statistical Science, Academia Sinica},
 title = {Network Influence Analysis},
 urldate = {2024-11-02},
 volume = {31},
 year = {2021}
}

@article{jin2024mixed,
  title={Mixed Membership Estimation for Social Networks},
  author={Jin, Jiashun and Ke, Zheng Tracy and Luo, Shengming},
  journal={Journal of Econometrics},
  volume={239},
  number={2},
  pages={105369},
  year={2024},
  publisher={Elsevier}
}

@article{lu2011link,
  title={Link Prediction in Complex Networks: A Survey},
  author={L{\"u}, Linyuan and Zhou, Tao},
  journal={Physica A: statistical mechanics and its applications},
  volume={390},
  number={6},
  pages={1150--1170},
  year={2011},
  publisher={Elsevier}
}

@article{albert2002statistical,
  title={Statistical Mechanics of Complex Networks},
  author={Albert, R{\'e}ka and Barab{\'a}si, Albert-L{\'a}szl{\'o}},
  journal={Reviews of Modern Physics},
  volume={74},
  number={1},
  pages={47},
  year={2002},
  publisher={APS}
}

@article{stephen2009explaining,
  title={Explaining the Power-Law Degree Distribution in a Social Commerce Network},
  author={Stephen, Andrew T and Toubia, Olivier},
  journal={Social Networks},
  volume={31},
  number={4},
  pages={262--270},
  year={2009},
  publisher={Elsevier}
}

@article{sengupta2018block,
  title={A Block Model for Node Popularity in Networks With Community Structure},
  author={Sengupta, Srijan and Chen, Yuguo},
  journal={Journal of the Royal Statistical Society Series B: Statistical Methodology},
  volume={80},
  number={2},
  pages={365--386},
  year={2018},
  publisher={Oxford University Press}
}

@article{krivitsky2009representing,
  title={Representing Degree Distributions, Clustering, and Homophily in Social Networks With Latent Cluster Random Effects Models},
  author={Krivitsky, Pavel N and Handcock, Mark S and Raftery, Adrian E and Hoff, Peter D},
  journal={Social Networks},
  volume={31},
  number={3},
  pages={204--213},
  year={2009},
  publisher={Elsevier}
}

@article{hoff2003,
  title={Random Effects Models for Network Data},
  author={Hoff, Peter D},
  journal={Dynamic Social Network Modeling and Analysis: Workshop Summary and Papers},
publisher={The National Academies Press},
  year={2003}
}

@article{wu2024bipartite,
  title={Bipartite Network Influence Analysis of a Two-Mode Network},
  author={Wu, Yujia and Lan, Wei and Fan, Xinyan and Fang, Kuangnan},
  journal={Journal of Econometrics},
  volume={239},
  number={2},
  pages={105562},
  year={2024},
  publisher={Elsevier}
}

@article{wu2022inward,
  title={Inward and Outward Network Influence Analysis},
  author={Wu, Yujia and Lan, Wei and Zou, Tao and Tsai, Chih-Ling},
  journal={Journal of Business \& Economic Statistics},
  volume={40},
  number={4},
  pages={1617--1628},
  year={2022},
  publisher={Taylor \& Francis}
}

@article{lee2004asymptotic,
  title={Asymptotic Distributions of Quasi-Maximum Likelihood Estimators for Spatial Autoregressive Models},
  author={Lee, Lung-Fei},
  journal={Econometrica},
  volume={72},
  number={6},
  pages={1899--1925},
  year={2004},
  publisher={Wiley Online Library}
}

@article{GAO2024121634,
title = {Citation Counts Prediction of Statistical Publications Based on Multi-Layer Academic Networks via Neural Network Model},
journal = {Expert Systems with Applications},
volume = {238},
pages = {121634},
year = {2024},
issn = {0957-4174},
doi = {https://doi.org/10.1016/j.eswa.2023.121634},
url = {https://www.sciencedirect.com/science/article/pii/S095741742302136X},
author = {Tianchen Gao and Jingyuan Liu and Rui Pan and Hansheng Wang},
}

@article{link2007,
author = {Liben-Nowell, David and Kleinberg, Jon},
title = {The Link-Prediction Problem for Social Networks},
journal = {Journal of the American Society for Information Science and Technology},
volume = {58},
number = {7},
pages = {1019-1031},
doi = {https://doi.org/10.1002/asi.20591},
url = {https://onlinelibrary.wiley.com/doi/abs/10.1002/asi.20591},
eprint = {https://onlinelibrary.wiley.com/doi/pdf/10.1002/asi.20591},
year = {2007}
}

@article{li2020network,
  title={Network Cross-Validation by Edge Sampling},
  author={Li, Tianxi and Levina, Elizaveta and Zhu, Ji},
  journal={Biometrika},
  volume={107},
  number={2},
  pages={257--276},
  year={2020},
  publisher={Oxford University Press}
}

@article{Ma2020UniversalLSM,
  title   = {Universal Latent Space Model Fitting for Large Networks With Edge Covariates},
  author  = {Ma, Zhuang and Ma, Zongming and Yuan, Hongsong},
  journal = {Journal of Machine Learning Research},
  year    = {2020},
  volume  = {21},
  number  = {4},
  pages   = {1--67},
  url     = {https://jmlr.org/papers/v21/17-470.html}
}

@article{Li2023InferenceLSM,
  title   = {Statistical Inference on Latent Space Models for Network Data},
  author  = {Li, Jinming and Wu, Shihao and Cui, Chengyu and Xu, Gongjun and Zhu, Ji},
  journal = {arXiv preprint arXiv:2312.06605},
  year    = {2023},
  url     = {https://arxiv.org/abs/2312.06605}
}

@article{Athreya2018RDPGSurvey,
  title   = {Statistical Inference on Random Dot Product Graphs: A Survey},
  author  = {Athreya, Avanti and Fishkind, Donniell E. and Tang, Minh and Priebe, Carey E. and Park, Youngser and Vogelstein, Joshua T. and Levin, Keith and Lyzinski, Vince and Qin, Yichen and Sussman, Daniel L.},
  journal = {Journal of Machine Learning Research},
  year    = {2018},
  volume  = {18},
  pages   = {1--92},
  url     = {http://jmlr.org/papers/v18/17-448.html}
}

@article{li2025high,
    author = {Li, Jinming and Xu, Gongjun and Zhu, Ji},
    title = {High-Dimensional Factor Analysis for Network-Linked Data},
    journal = {Biometrika},
    volume = {112},
    number = {4},
    pages = {asaf012},
    year = {2025},
    month = {08},
}

@article{yan2016asymptotics,
  title={Asymptotics in Directed Exponential Random Graph Models With an Increasing Bi-degree Sequence},
  author={Yan, Ting and Leng, Chenlei and Zhu, Ji},
  journal={The Annals of Statistics},
  volume={44},
  number={1},
  pages={31--57},
  year={2016},
  publisher={Institute of Mathematical Statistics}
}

@article{wasserman1994social,
  title={Social Network Analysis: Methods and Applications},
  author={Wasserman, Stanley},
  journal={The Press Syndicate of the University of Cambridge},
  year={1994}
}
\end{center}

\end{document}